\documentclass[journal,comsoc,twocolumn]{IEEEtran}
\usepackage{stfloats}
\usepackage{amsthm}
\usepackage{amsmath}
\usepackage{bbm}
\usepackage{mathrsfs}
\usepackage[numbers,sort&compress]{natbib}
\usepackage{bm,epsfig,amsthm,url}
\usepackage{indentfirst} 
\usepackage{amssymb}
\usepackage{amsfonts}
\usepackage{epstopdf}
\usepackage{algorithm}
\usepackage{algorithmic}
\usepackage[caption=false,font=footnotesize]{subfig}
\usepackage{xcolor}
\usepackage{makecell}
\usepackage{mathtools}

\usepackage{hyperref}
\usepackage{balance}
\usepackage{multirow}

\def\BibTeX{{\rm B\kern-.05em{\sc i\kern-.025em b}\kern-.08em
    T\kern-.1667em\lower.7ex\hbox{E}\kern-.125emX}}
\usepackage{graphicx}
\usepackage{epstopdf}

\begin{document}
\title{Out-of-Band Modality Synergy Based Multi-User Beam Prediction and Proactive BS Selection with Zero Pilot Overhead}
\author{Kehui Li, Binggui Zhou, Jiajia Guo, Feifei Gao, Guanghua Yang, and Shaodan Ma

\thanks{This work was supported in part by the National Key R\&D Program of China under Grant 2024YFE0200700, in part by National Natural Science Foundation of China under Grants 62171201, 62261160650, and 62401640, in part by the Science and Technology Development Fund, Macau SAR under Grants 0087/2022/AFJ, 0114/2025/AMJ, 0020/2025/RIB1, and 001/2024/SKL, in part by the Research Committee of University of Macau under Grant MYRG-GRG2025-00143-IOTSC, and in part by the Guangdong Basic and Applied Basic Research Foundation under Grant 2023A1515110732. 
(\emph{Corresponding Authors: Binggui Zhou, Guanghua Yang}.)}

\thanks{
Kehui Li is with the School of Intelligent Systems Science and Engineering, Jinan University, Zhuhai 519070, China; and also with the State Key Laboratory of Internet of Things for Smart City and the Department of Electrical and Computer Engineering, University of Macau, Macao 999078, China (e-mails: yc47997@um.edu.mo).}

\thanks{
Binggui Zhou is with the Department of Electrical and Electronic Engineering, Imperial College London, SW7 2AZ London, U.K. (e-mail: binggui.zhou@imperial.ac.uk).}

\thanks{
Jiajia Guo and Shaodan Ma are with the State Key Laboratory of Internet of Things for Smart City and the Department of Electrical and Computer Engineering, University of Macau, Macao 999078, China (e-mails: jiajiaguo@um.edu.mo; shaodanma@um.edu.mo).}

\thanks{
Feifei Gao is with the Institute for Artificial Intelligence, Tsinghua University (THUAI), State Key Lab of Intelligent Technologies and Systems, Tsinghua University, Beijing National Research Center for Information Science and Technology (BNRist), and Department of Automation, Tsinghua University, Beijing 100084, China (email: feifeigao@ieee.org).}

\thanks{
Guanghua Yang is with the School of Intelligent Systems Science and Engineering and the GBA and B\&R International Joint Research Center for Smart Logistics, Jinan University, Zhuhai 519070, China (e-mail: ghyang@jnu.edu.cn).}
}
\maketitle

\begin{abstract}
Multi-user millimeter-wave communication relies on narrow beams and dense cell deployments to ensure reliable connectivity. However, tracking optimal beams for multiple mobile users across multiple base stations (BSs) results in significant signaling overhead. Recent works have explored the capability of out-of-band (OOB) modalities in obtaining spatial characteristics of wireless channels and reducing pilot overhead in single-BS single-user/multi-user systems. However, applying OOB modalities for multi-BS selection towards dense cell deployments leads to high coordination overhead, i.e, excessive computing overhead and high latency in data exchange. How to leverage OOB modalities to eliminate pilot overhead and achieve efficient multi-BS coordination in multi-BS systems remains largely unexplored. In this paper, we propose a novel OOB modality synergy (OMS) based mobility management scheme to realize multi-user beam prediction and proactive BS selection by synergizing two OOB modalities, i.e., vision and location. Specifically, mobile users are initially identified via spatial alignment of visual sensing and location feedback, and then tracked according to the temporal correlation in image sequence. Subsequently, a binary encoding map based gain and beam prediction network (BEM-GBPN) is designed to predict beamforming gains and optimal beams for mobile users at each BS, such that a central unit can control the BSs to perform user handoff and beam switching. Simulation results indicate that the proposed OMS-based mobility management scheme enhances beam prediction and BS selection accuracy, achieves up to $86\%$ of the optimal transmission rate with zero pilot overhead, and significantly improves multi-BS coordination efficiency compared to existing methods.
\end{abstract}
\begin{IEEEkeywords}
Out-of-band modality sensing, multi-user, beam prediction, proactive handoff, mobility management, deep learning.
\vspace{-6mm}
\end{IEEEkeywords}
\section{Introduction}
Millimeter-wave (mmWave) and sub-terahertz communications are emerging as pivotal technologies for both current and future wireless networks. These frequency bands offer extensive bandwidths to meet the high data rate requirements of various advanced applications, including wireless virtual/augmented reality (VR/AR) and autonomous driving systems \cite{xue2024surveya}. However, transmission at these frequency bands is highly susceptible to atmospheric loss and blockage attenuation, leading to a significant decline in the received signal-to-noise ratio (SNR). To overcome these challenges, highly directional antenna arrays are used to focus the wireless signaling energy into a narrow beam directed toward users. Nevertheless, identifying the optimal beam direction through exhaustive search across all possible angles incurs substantial pilot overhead. Furthermore, the performance of communications is prone to degrade due to the limited diffraction capability and narrow beamwidth of mmWave, which complicates the deployment and reliability of mmWave communications systems \cite{yi2024beam}. To address these challenges, dense cell deployments \cite{wang2018millimeter} can be utilized to ensure reliable connectivity. Thus, effectively selecting optimal beams across multiple base stations (BSs) becomes the primary challenge in mmWave communications.

Since radio-frequency (RF) signal propagation in high-frequency bands becomes increasingly dependent on the spatial characteristics of the physical environment, and the communication range in mmWave and terahertz bands approaches the field of view of humans or cameras, sensing technologies that capture the surrounding environment through various OOB modalities (e.g., RGB camera, depth camera, LiDAR, and global positioning system (GPS) trackers) has received considerable attention these days \cite{kim2024rolea}. These modalities can be leveraged to improve environmental awareness and acquire more diverse communication-related features compared to traditional RF sensing, boosting the communication performance and reducing the pilot overhead in various communication tasks \cite{roy2023going}, such as beam training, channel estimation, and BS selection, etc. 
\subsection{Beam Training}
Beam management in mmWave communications systems has two key challenges: improving beam training performance and reducing pilot overhead. Traditional solutions can be roughly divided into three categories: (i) Constructing adaptive beam codebooks to reduce beam search time and save spectrum resources \cite{qi2020hierarchicala}; (ii) Designing beam tracking techniques to mitigate frequent beam sweeping \cite{jayaprakasam2017robust}, or beam prediction techniques to enable proactive beam switching \cite{li2023machine}; (iii) Exploiting channel state information (CSI) or signal echoes to estimate the location or trajectory of mobile users (MUs), thereby determining the optimal beam direction directly \cite{chen2023accurate, zhao2025interferencerobust, zhao2025nearfield}. However, these approaches inevitably utilize pilot signals to perceive MUs, which introduce significant pilot overhead increasing with the scale of transmit antenna array and the number of users, whereas visual modality can provide OOB sensing capability at the same cost, regardless of the scale of transmit antenna arrays and user density. In \cite{xu2023computer}, the camera on a vehicle was utilized to perceive the surrounding environment for beam alignment, which eliminated the pilot overhead. In \cite{jiang2022computer}, the visual data captured by cameras at the BS was utilized to enable a fast beam tracking and refining process in a prototype system. In addition, the location data of both the receiver vehicle and its neighboring vehicles were exploited in \cite{wang2019mmwave} to capture situational awareness, thereby reducing beam search overhead.
\begin{figure*}[tp] 
    \centering
    \includegraphics[scale=0.5]{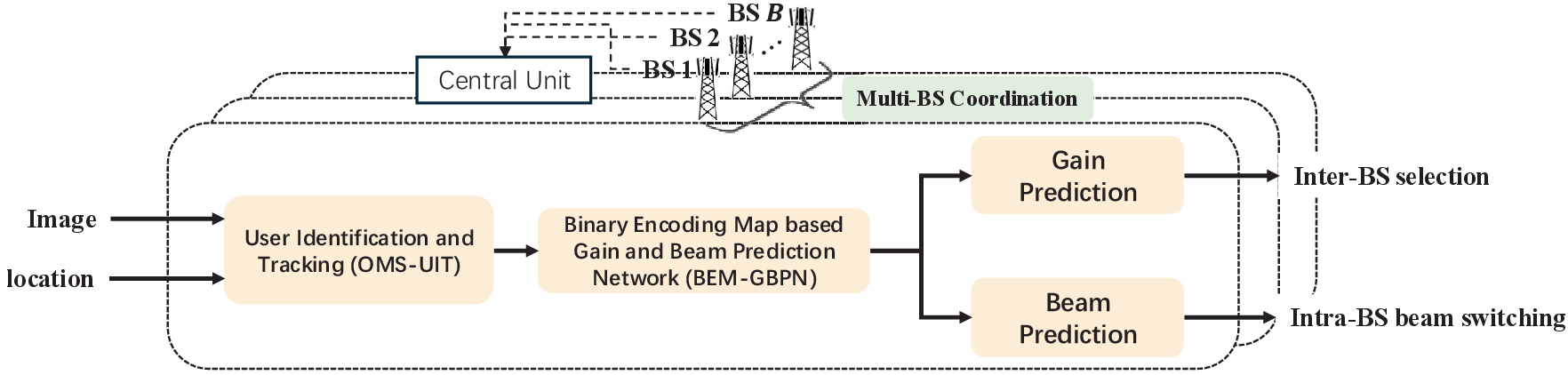}
    \caption{The framework diagram of the proposed out-of-band modality synergy based mobility management scheme.}
    \label{Whole_diagram}
    \vspace{-0.5cm}
\end{figure*}

Although these studies have shown that utilizing OOB modalities is able to enhance beam training performance while reducing the overhead in single-user systems, how multiple OOB modalities can be synergized to identify and distinguish users in multi-user communication systems is still challenging and remains uninvestigated. To cope with these challenges, \cite{ahn2022machinea} assumed all visually detected objects were users, which is not feasible in a practical scenario. The works \cite{xu2023multiusera, charan2024user, depinho2020visionaideda} utilized beam sweeping to identify and distinguish users in visual data. However, continual beam sweeping synchronized with image acquisition introduces substantial pilot overhead, which contradicts the original intention of using OOB modality. Therefore, how to fully reap the benefits from OOB sensing capability to reduce communication overhead in multi-user systems deserves further investigation. 
\subsection{Base Station Selection}
Line-of-sight (LOS) link blockage and long-distance path loss are significant challenges in high-frequency band wireless communications \cite{scarborough2018beamforming}. To address these issues, the ultra dense network (UDN) has emerged as a promising solution, where a number of small BSs are densely deployed within the coverage area of a macro BS. To fully leverage the advantages of the UDN, it is essential to associate MUs with nearby BSs to ensure high quality-of-service (QoS). Traditionally, the BS selection is supported by utilizing pilot signals to measure the reference signal received power (RSRP) at the user side. When the attenuation/enhancement of the RSRP is detected, the handoff process is initiated \cite{giordani2016multiconnectivity}. However, these reactive approaches only respond to worse communication links, where connectivity is threatened. To ensure reliable connectivity, the authors in \cite{mismar2023deepa, huang2020machine} utilized deep neural networks (DNNs) and time series data generated from pilot sensing to predict the quality of communication and enable proactive handoff. However, the signal-to-interference-plus-noise ratio can suddenly drop up to $35$dB if the communication link changes from LoS to non-line-of-sight (NLoS) condition, e.g., occlusion by the scatterer. Consequently, relying solely on the RSRP or CSI lacks information about the spatial distributions of the surrounding non-user scatterers, leading to a degradation in BS selection accuracy and user throughput. In contrast, \cite{lin2024multicameraa,charan2021visionaideda,liu2023visionaideda} explored how camera image sequences can assist blockage prediction and guide BS selection proactively, while \cite{koda2020handover, ahn2024sensinga} utilized visual sensing to directly predict communication rate and enable proactive handoff to the optimal link. 

Although these approaches utilize OOB modality sensing to increase environmental awareness and enable accurate proactive BS selection, they still rely on frequent pilot measurements to differentiate users and scatterers, which thus introduces significant sensing latency and pilot overhead. Furthermore, the raw data of OOB modality is inherently enormous, such that relying on a central unit to process images from all BSs poses a significant coordination overhead \cite{xu2023multiusera, lin2024multicameraa, ahn2024sensinga}. Hence, a framework that can fully leverage the sensing advantages of OOB modality and achieve efficient coordination among BSs is desired.
\subsection{Contributions}
In summary, the primary challenges of integrating OOB modality sensing into multi-BS and multi-user communication systems for mobility management are twofold: (i) a single OOB sensor struggles to distinguish between users and scatterers, while combining it with wireless pilot sensing will bring additional pilot overhead. (ii) Excessive OOB modality data from multiple BSs imposes an overwhelming coordination overhead, including a huge computational burden on the central unit and significant latency in data exchange among BSs. Fortunately, since user location feedback can also be exploited to identify users in visual data and eliminate pilot overhead, we propose a novel mobility management scheme that relies on the synergy of OOB modalities, i.e., vision and location, to address these challenges. This OOB modality synergy (OMS) based mobility management scheme is designed to operate locally within each BS utilizing its own computational capacity, and only the prediction results, which occupy minimal data size (a few bytes), are shared with the central unit for coordination. Overall, the proposed OMS-based mobility management scheme can achieve zero pilot overhead and efficient multi-BS coordination, while enhancing beam prediction and BS selection accuracy in multi-user and dense cell deployment systems. The whole structure of the OMS-based mobility management scheme is shown in Fig. \ref{Whole_diagram}. The main contributions are as follows:
\begin{figure*}[!tbp]
\begin{equation}\label{eq1}
y_u =
(\boldsymbol{h}_{b_u,u})^H
\boldsymbol{F}_{b_u}^{RF} [\boldsymbol{F}_{b_u}^{BB}]_{:,u} s_u +  \underbrace{\sum_{u'\in\mathcal{U}_{b_u}\setminus{u}} (\boldsymbol{h}_{b_u,u})^H \boldsymbol{F}_{b_u}^{RF} [\boldsymbol{F}_{b_u}^{BB}]_{:,u'} s_{u'}}_{\text{intra-BS interference}} + \underbrace{\sum_{b_{u'}\neq b_u}\sum_{u'\in\mathcal{U}_{b_{u'}}} (\boldsymbol{h}_{b_{u'},u})^H \boldsymbol{F}_{b_{u'}}^{RF} [\boldsymbol{F}_{b_{u'}}^{BB}]_{:,u'} s_{u'}}_{\text{inter-BS interference}}
+ n_u.    
\end{equation}
\vspace{-0.7cm}
\end{figure*}
\begin{itemize}
  \item We propose a novel sensing algorithm, i.e., OOB modality synergy based user identification and tracking (OMS-UIT), to differentiate users and scatterers, and track MUs without any pilot overhead. Unlike existing OOB modality-aided communications systems that require additional wireless sensing to identify users, the OMS-UIT algorithm eliminates RF sensing overhead while fully reaping the benefits of OOB modalities in high-accuracy and efficient sensing. 
  \item We design a binary encoding map based gain and beam prediction network (BEM-GBPN) to predict beamforming gains and optimal beams for MUs at each BS. The consecutive OOB data are first transformed into a binary encoding map (BEM) sequence based on the results from the OMS-UIT algorithm. Subsequently, a sophisticated feature extractor is proposed to capture the spatio-temporal characteristics of the BEM sequence. Then, a mixture-of-experts (MoE) structure is proposed to generate various semantic features and adaptively combine these semantic features, thereby improving the gain and beam prediction accuracy.
  \item We further propose a proactive BS selection and beam switching approach for multi-user multi-BS mmWave networks. First, BSs leverage BEM-GBPN to locally predict beamforming gains and optimal beams for detected MUs, and transmit these predictions to a central unit, which contains compact data (e.g., user identities and predicted gains). The central unit then proactively assigns each MU to the BS with the highest predicted beamforming gain, and the selected BS configures its analog beamformer based on the optimal beam prediction. This approach can enable a seamless handoff and reliable connectivity for multi-user, while achieving efficient multi-BS coordination.
  \item Finally, we construct an urban simulation platform with multiple dynamic vehicles to evaluate the OMS-based mobility management scheme by generating synchronized OOB data and wireless channels. Experimental results demonstrate that the BEM-GBPN significantly outperforms existing methods in both beamforming gain and beam prediction accuracy. Additionally, the proposed proactive BS selection and beam switching approach enables a better transmission rate than the traditional reactive scheme, while eliminating pilot overhead and reducing multi-BS coordination overhead.
  \end{itemize}

The remainder of this paper is organized as follows. Section II introduces the system model and the problem formulation. Section III proposes the OMS-based mobility management scheme. Section IV provides the simulation setup, numerical results, and detailed discussions. Finally, the conclusions are given in Section V.

\emph{Notation:} $\boldsymbol{A}$ is a matrix or tensor; $\bm{\mathcal{A}}$ is a set; $\boldsymbol{a}$ is a vector; $a$ is a scalar; $[\boldsymbol{A}]_{i, j}$ is the element of the $i$-th row and the $j$-th column in $\boldsymbol{A}$; $[\boldsymbol{A}]_{i, :}$ and $[\boldsymbol{A}]_{:, j}$ are the $i$-th row and the $j$-th column of $\boldsymbol{A}$ respectively; $\bm{\mathcal{A}} \setminus \bm{\mathcal{A}}'$ is the set $\bm{\mathcal{A}}$ that excludes all elements from $\bm{\mathcal{A}}'$.
\begin{figure*}[tp] 
    \centering
    \includegraphics[scale=0.6]{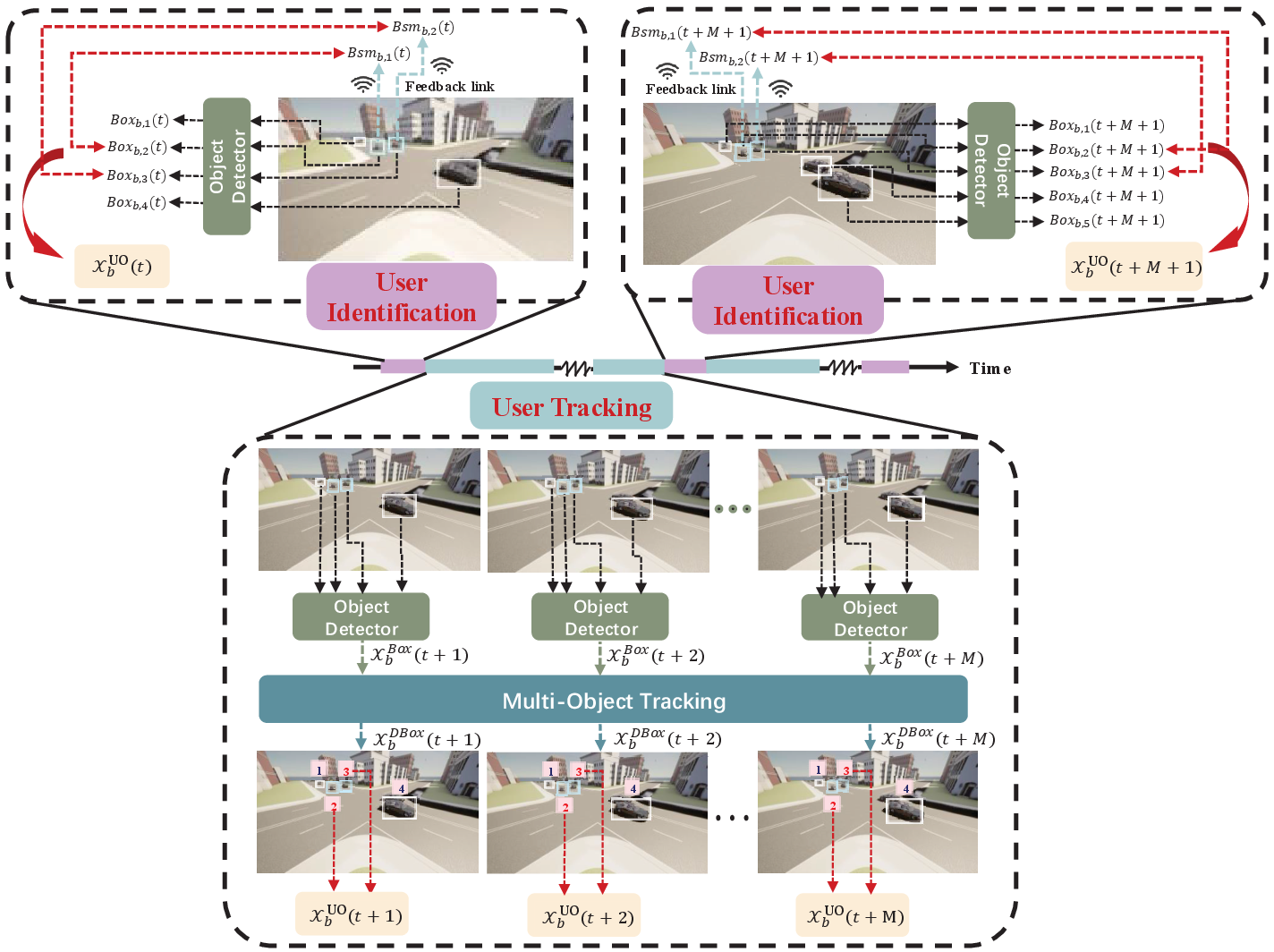}
    \caption{The diagram of the OMS-based user identification and tracking algorithm. First, the BSM data fed back from users is utilized to identify user objects in the image. Subsequently, the visual MOT algorithm tracks these user objects in real-time. Aperiodic BSM feedback from users is used to correct the tracking results to ensure robust user identification and tracking.}
    \label{UITB_OMS}
    \vspace{-0.5cm}
\end{figure*}
\vspace{-0.2cm}
\section{System Model and Problem Formulation}
\subsection{System Model}
We consider a downlink mmWave communications system, and there are $B$ BSs with the hybrid beamforming architecture and $U$ MUs. The $u$-th MU, $\forall u \in \mathcal{U}=\{1,2,..., U\}$, is equipped with an omnidirectional single-antenna. The $b$-th BS, $\forall b \in \mathcal{B}=\{1,2,..., B\}$ is equipped with an RGB camera and a uniform linear array (ULA) of $N_{T}$ antennas and $N_{RF}$ RF chains. At each time slot, one MU can only connect to one BS while one BS can serve multiple MUs. Denote $U_b \leq N_{RF}$ as the number of MUs served by the $b$-th BS. 

During the downlink signal transmission, the received signal of the $u$-th MU, denoted by $y_u$, can be expressed as (\ref{eq1}) at the top of this page, where $b_u \in \bm{\mathcal{B}}$ and $\boldsymbol{h}_{b_u,u} \in 
\mathbb{C}^{N_{T} \times 1}$ represent the BS that the $u$-th MU is connected to and the channel vector between the $b_u$-th BS and the $u$-th MU. $\boldsymbol{F}^{RF}_{b_u} \in \mathbb{C}^{N_{T} \times N_{RF}} $, $\boldsymbol{F}^{BB}_{b_u} \in \mathbb{C}^{N_{RF} \times U}$, and $s_u$ represent the analog beamformer, the digital beamformer, and the transmitted signal at the $b_u$-th BS, respectively, where the transmitted signal is subject to the power constraint, i.e., $\mathbb{E}\{\left|s_u \right|^2 \} = 1$. In addition, the received signal $y_u$ is impaired by intra-BS interference when $b_{u'} = b_u$, and inter-BS interference when $b_{u'} \neq b_u$. In the intra-BS interference term, $\mathcal{U}_{b_u}$ denotes the set of MUs served by the $b_u$-th BS. In the inter-BS interference term, $\boldsymbol{F}^{\mathrm{RF}}_{b_{u’}} \in \mathbb{C}^{N_T \times N_{\mathrm{RF}}}$, $\boldsymbol{F}^{\mathrm{BB}}_{b_{u’}} \in \mathbb{C}^{N_{\mathrm{RF}} \times U}$, and $\mathcal{U}_{b_{u’}}$ represent the analog beamformer, the digital beamformer, and the set of MUs served by the $b_{u’}$-th BS, respectively. The additive white Gaussian noise (AWGN) at the $u$-th user is denoted as $n_u\in \mathcal{CN}(0,\sigma^2)$. 

According to the Saleh-Valenzuela channel model \cite{heath2016overview}, the channel vector $\boldsymbol{h}_{b,u}$ can be expressed as
\begin{equation}\label{eq2}
    \boldsymbol{h}_{b,u}=\sqrt{\frac{1}{L_{b,u}}} \sum_{l=1}^{L_{b,u}} \lambda_l \boldsymbol{\alpha}(\theta_l),
\end{equation}
\noindent where $L_{b,u}$ denotes the number of paths, and $\lambda_l$ and $\theta_l$ represent the channel gain and the angle-of-departure (AoD) of the $l$-th path, respectively. In addition, the channel steering vector in \eqref{eq2} is defined as
\begin{equation}\label{eq3}
    \boldsymbol{\alpha}(\theta_l)=\left[1, e^{j 2\pi \frac{d_a}{\lambda_a} \cos{(\theta_l)}}, \cdots, e^{j2\pi(N_{T}-1) \frac{d_a}{\lambda_a} \cos{(\theta_l)}}\right]^T,
\end{equation}
\noindent where $d_a$ is the distance between adjacent antennas and $\lambda_a$ is the carrier wavelength. 

\subsection{Problem Formulation}
Our objective is to maximize the achievable sum-rate of $U$ MUs by first selecting the optimal BS for each MU, and then configuring the analog and digital beamformers of each BS. The whole process can be formulated as
\begin{subequations}\label{eq4}
\begin{align}
     \underset{b_u,\boldsymbol{F}^{RF}_{b_u}, \boldsymbol{F}^{BB}_{b_u}} {\max}& \sum_{u\in\mathcal{U}} R_u  \\
     \text {s.t. }\ \ \ \ \ 
     & \sum_{u \in \mathcal{U}} \mathbbm{1} \left(b_u=b\right) \leq N_{RF}, \quad \forall b \in \bm{\mathcal{B}},  \label{eq4b}  \\ 
     & \sum_{b \in \mathcal{B}} \mathbbm{1} \left(b=b_u\right)=1, \quad \forall u \in \bm{\mathcal{U}}, \label{eq4c}   \\  
     & \left\| \boldsymbol{F}^{RF}_{b_u} \boldsymbol{F}^{BB}_{b_u} \right\|_2=U_{b_u}, \label{eq4d} \\
     & [\boldsymbol{F}^{RF}_{b_u}]_{:,u} \in \bm{\mathcal{F}},  
\end{align}
\end{subequations}
\noindent where 
\begin{equation} \label{eq5}
\begin{split}
    R_u=\log _2\left(1+\frac{\left|(\boldsymbol{h}_{b_u,u})^H \boldsymbol{F}^{RF}_{b_u} [\boldsymbol{F}^{BB}_{b_u}]_{:,u}\right|^2}{\sum_{u' \neq u} \left|(\boldsymbol{h}_{b_{u'},u})^H \boldsymbol{F}^{RF}_{b_{u'}} [\boldsymbol{F}^{BB}_{b_{u'}}]_{:,u'}\right|^2+\sigma^2}\right)
\end{split}
\end{equation}
is the achievable rate of the $u$-th MU. The indicator function $\mathbbm{1} \left(\mathcal{C}\right)$ is set as $1$ when the condition $\mathcal{C}$ is true. The condition (\ref{eq4b}) constrains the maximum number of serving MUs at each BS, and each MU can only connect to one BS as stated in (\ref{eq4c}). In addition, $\bm{\mathcal{F}}=\{\boldsymbol{f}_1,\boldsymbol{f}_2,...,\boldsymbol{f}_{N_{T}}\}$ represents the analog beamforming codebook adopted by the BS and each beam in the codebook can be expressed as
\begin{equation}\label{eq6}
    \boldsymbol{f}_n=\frac{1}{\sqrt{N_{T}}}\left[1, e^{j 2\pi \frac{d_a}{\lambda_a}n \frac{1}{N_{T}}}, \cdots, e^{j2\pi \frac{d_a}{\lambda_a} n \frac{N_T-1}{N_{T}}}\right]^T. 
\end{equation}

The problem stated in (\ref{eq4}) necessitates a joint optimization of BS selection and beam training. Traditional wireless sensing based approaches depend on pilots to perceive the MUs' location, resulting in significant pilot overhead, especially when the number of MUs is large or dynamic beam tracking is required. Although leveraging OOB modality can reduce the pilot overhead, excessive OOB modality data will introduce heavy coordination overhead in multi-BS systems. In the following section, we propose the OMS-based mobility management scheme to eliminate the pilot overhead and achieve efficient multi-BS coordination.
\section{OMS-based Mobility Management Scheme}
\label{OMS_UIT}
In this section, we introduce the OMS-based mobility management scheme, which comprises three essential components: the OMS-UIT algorithm, the BEM-GBPN network, and the proactive BS selection and beam switching approach. Specifically, MUs can be first identified and tracked utilizing two OOB modalities, i.e., vision and location, without any pilot overhead. Subsequently, the BEM-GBPN is proposed to predict the beamforming gains and optimal beams for MUs at each BS. Finally, the prediction results from all BSs are transmitted to a central unit, which proactively assigns each MU to the BS with the highest predicted beamforming gain, and configures its analog beamformer based on the optimal
beam prediction.
\subsection{Users Identification and Tracking}
\label{OMS_UI}
In vision-aided communications systems, user objects (UOs) must be identified first in images. To support user identification, environmental objects that are potential user objects (PUOs) should be extracted from images using object detection techniques. For simplicity, we assume all PUOs are vehicles in this paper \footnote{The OMS-UIT algorithm can be readily extended to broader communication scenarios, including those involving pedestrians or other mobile entities, simply by training the object detector on more diverse datasets. }. To detect all vehicles in an image efficiently, we employ a state-of-the-art (SOTA) object detector in general computer vision field, i.e., the You Only Look Once v5 (YOLOv5) \cite{YOLOv5}. However, how to identify the UO vehicles from all the detected PUO vehicles is still challenging, even when we can detect all PUO vehicles correctly via the YOLOv5 visual object detector.
\begin{figure}[tp!]
    \centering
    \includegraphics[scale=0.7]{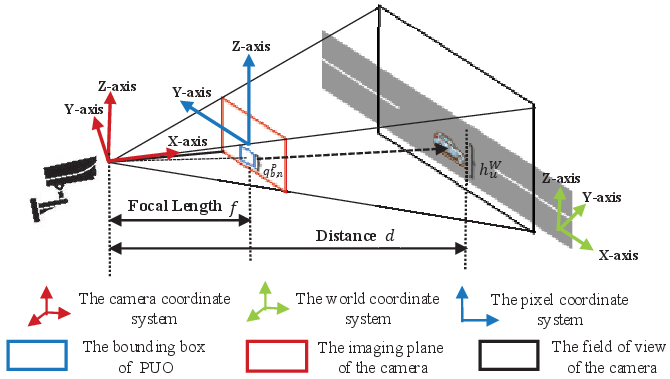}
    \caption{The illustration of the camera imaging principle and the relations between different coordinate systems. }
    \label{Imaging_principle}
    \vspace{-0.5cm}
\end{figure}
\begin{figure*}[!bh]
    \begin{equation}\label{eq7}
    \begin{array}{*{20}{c}}
    {\left[ {\begin{array}{*{20}{l}}
    {x_{b,u}^{\rm{C}}}\\
    {y_{b,u}^C}\\
    {z_{b,u}^C}
    \end{array}} \right] = \left[ {\begin{array}{*{20}{c}}
    1&0&0\\
    0&{\cos \beta _b^W}&{\sin \beta _b^W}\\
    0&{ - \sin \beta _b^W}&{\cos \beta _b^W}
    \end{array}} \right]} 
    {\left[ {\begin{array}{*{20}{c}}
    {\cos \alpha _b^W}&0&{ - \sin \alpha _b^W}\\
    0&1&0\\
    {\sin \alpha _b^W}&0&{\cos \alpha _b^W}
    \end{array}} \right]\left[ {\begin{array}{*{20}{c}}
    {x_u^W}\\
    {y_u^W}\\
    {z_u^W}
    \end{array}} \right] + \left[ {\begin{array}{*{20}{c}}
    {X_b^W}\\
    {Y_b^W}\\
    {Z_b^W}
    \end{array}} \right]}
    \end{array}.
    \end{equation}
\end{figure*}

Traditionally, wireless sensing data, e.g., the RSRP data obtained from beam sweeping \cite{charan2024user} and the CSI obtained from channel estimation \cite{guo2025prompt}, can be utilized to identify UOs via aligning with visual sensing data. However, these wireless sensing approaches will bring in unaffordable pilot overhead, especially in dynamic multi-user scenarios, and aligning wireless sensing with visual perception in many time slots introduces significant sensing delays. Fortunately, modern vehicles equipped with numerous sensors for driving assistance, e.g., GPS and Beidou Navigation Satellite System (BDS), are becoming increasingly common \cite{salehi2022deep}. Besides, the society of automotive engineers (SAE) standard J2735 defines syntax and semantics of Vehicle-to-Everything (V2X) messages, in which the basic safety message (BSM) conveys core state information about the sending vehicle, e.g., its identity, position, and size \cite{BSM_intro}. This information can be utilized to identify UOs while eliminating the pilot overhead associated with wireless sensing. However, in practical applications, leveraging the BSM to track UOs significantly affects tracking robustness due to the low positioning frequency of sensors and the unstable feedback link of the BSM. In contrast, image sequences of visual modality can be quickly and stably obtained at each BS. Therefore, to exploit the complementary advantages of these two modalities, and enable efficient and robust user identification and tracking, we propose the OMS-UIT algorithm as shown in Fig. \ref{UITB_OMS}. Specifically, user identification is first achieved by combining the BSM feedback of UOs and information about visually detected PUOs. Then, a visual multi-object tracking (MOT) technique is employed to track the identified UOs when the BSM feedback is absent. Once a new BSM arrives, the user identification process can be performed again. As such, the OMS-UIT algorithm is able to eliminate the overhead of wireless sensing and fully leverage the information from OOB modalities. In the following subsections, we elaborate on the details of the proposed OMS-UIT algorithm.
\subsubsection{User Identification}
Given an input image, the visual object detector outputs the bounding box (BBox) set, in which each bounding box is defined as $Box_{b,n} = [u^{P}_{b,n}, v^P_{b,n}, e^P_{b,n}, q^P_{b,n}]$, where $(u^P_{b,n}, v^P_{b,n})$ represents the coordinates of the $n$-th PUO relative to the 2D pixel coordinate system (PCS) with an origin at the center of the $b$-th camera's image, and $e^P_{b,n}$ and $q^P_{b,n}$ are the width and height of the $n$-th PUO's bounding box. Moreover, the BSM is defined as $Bsm'_{u}=[x^W_{u}, y^W_{u}, z^W_{u}, w^W_{u}, h^W_{u}]$, in which $( x^W_{u}, y^W_{u}, z^W_{u})$ represents the coordinates of the $u$-th UO relative to the 3D world coordinate system (WCS) with its origin at a specific point in the physical environment, and $w^W_{u}$ and $h^W_{u}$ are physical width and height of the $u$-th UO, respectively. The relation between each coordinate system is shown in Fig. \ref{Imaging_principle}. Note that the coordinate systems must be identical for utilizing the BSM to identify the BBox of the $u$-th UO.

Specifically, the coordinates $\left(X^W_{b}, Y^W_{b}, Z^W_{b}\right)$, the azimuth angle $\alpha_b^W$, and the elevation angle $\beta_b^W$ of the $b$-th camera relative to the WCS are obtained first. Then, the coordinates $( x^C_{b,u}, y^C_{b,u}, z^C_{b,u} )$ of the $u$-th UO relative to the $b$-th camera coordinate system (CCS) can be obtained via rotation and shifting transformations as (\ref{eq7}) at the bottom of this page. According to the principle of camera imaging \cite{cv_calibration}, we can transform the coordinates $( x^C_{b,u}, y^C_{b,u}, z^C_{b,u} )$ from the CCS to $( y^P_{b,u}, z^P_{b,u})$ in the PCS as
\begin{equation}\label{8}
    \left[ {\begin{array}{*{20}{c}}
    {y^P_{b, u}}\\
    {z^P_{b, u}}\\
    1
    \end{array}} \right] \!=\! \left[ {\begin{array}{*{20}{c}}
    {u_0}&{\frac{1}{{dy}}}&0\\
    {v_0}&0&{\frac{1}{{dz}}}\\
    1&0&0
    \end{array}} \right]\!\left[ {\begin{array}{*{20}{c}}
    1&0&0&0\\
    0&{f}&0&0\\
    0&0&{f}&0
    \end{array}} \right]\!\left[ {\begin{array}{*{20}{c}}
    {{x^C_{b, u}}}\\
    {{y^C_{b, u}}}\\
    {{z^C_{b, u}}}\\
    1
    \end{array}} \right],
\end{equation}
where $(u_0,v_0)$ represents the pixel coordinates of the image's center, and $f$ is the focal length of the camera lens on the optic axis $X$. Intuitively, in the PCS, one pixel occupies $dy$ unit lengths in the $Y$-axis direction and $dz$ unit lengths in the $Z$-axis direction. 
\begin{figure}[tp]
    \centering
    \includegraphics[scale=0.55]{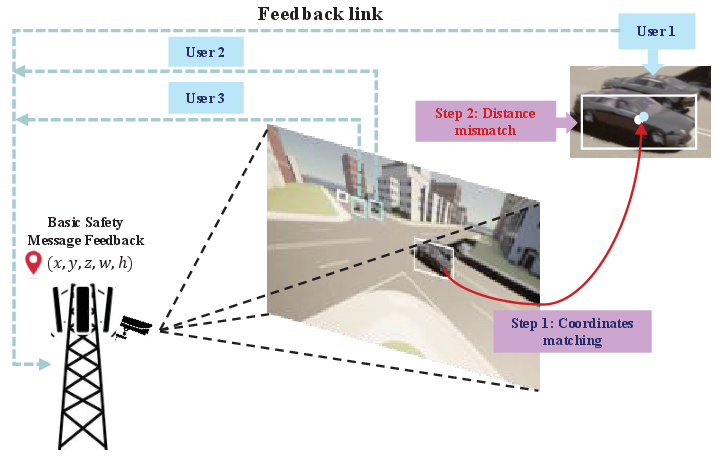}
    \caption{The illustration of the two-step matching based user identification process. A case of distance mismatch is shown on the top-right side, in which the user object is obstructed by the non-user object.}
    \label{UI_error}
    \vspace{-0.8cm}
\end{figure}

The commonly used monocular camera, which is adopted by the BS, however, lacks ability to sense in-depth information about environment. Relying solely on coordinate matching between $( y^P_{b,u}, z^P_{b,u})$ from the BSM and $( u^{P}_{b,n}, v^P_{b,n})$ from the BBox would lead to user identification error if other non-user objects are obstructing the UO at the same angle relative to the BS. For example, in Fig. \ref{UI_error}, user1 is occluded by a visually identical non-user vehicle. Mistakenly identifying this non-user vehicle as user1 can lead to subsequent visual tracking errors once the BSM feedback is lost, thereby severely degrading the communication performance. Thus, we propose a two-step matching approach to enable user identification via fully exploiting the spatial features of the image and the BSM. According to the imaging model of a small hole as previously shown in Fig. \ref{Imaging_principle}, the distance between the $n$-th PUO and the BS can be estimated as $\frac{h^W_u \times f}{q^{P}_{b,n}}$ with the physical height information $h^W_u$ of the $u$-th UO, and the real distance is $d^C_{b,u}=\sqrt{(x^C_{b,u})^2 + (y^C_{b,u})^2 + (z^C_{b,u})^2}$. Therefore, the $n$-th PUO from the BBox set is first tentatively identified as the $u$-th UO via coordinate matching, where the coordinate difference between the $u$-th UO and the $n$-th PUO in the image is smaller than the threshold $\gamma_1$ \footnote{$\gamma_1$ represents the maximum tolerance for pixel alignment error. It is empirically set to the sum of the width and height of the n-th PUO's BBox to accommodate potential detection inaccuracies. }. Subsequently, a distance matching process is performed to correct potential errors in the tentative identification by ensuring that the difference between the visually estimated distance and the actual distance between the UO and the BS is smaller than $\gamma_2$ \footnote{$\gamma_2$ denotes the distance estimation error tolerance. It is set to $3$ m to align with the typical highway lane width for reliable association within the same spatial lane in the exampled case, and can be adjusted accordingly in other cases. }. 

We denote the BBox set of all PUOs relative to the $b$-th BS at the $t$-th frame as $\boldsymbol{\mathcal{X}}^{Box}_b(t)=\left\{Box_{b,n}(t)|n=1,2,..., N^P_b(t), b \in \bm{\mathcal{B}}\right\}$ and the BSM set of all UOs as $\boldsymbol{\mathcal{X}}^{Bsm}_b(t)=\left\{Bsm_{b,u}(t)|u=1,2,..., N^U_b(t), b \in \bm{\mathcal{B}}\right\}$, where $Bsm_{b,u}=[y^P_{b,u}, z^P_{b,u}, d^C_{b,u}, h^W_{u}]$, and $N^P_b(t)$ and $N^U_b(t)$ denote the number of PUOs and UOs, respectively. The pseudo code of the two-step matching based user identification algorithm is shown in Algorithm 1.
\subsubsection{User Tracking}
After user identification, we can get the BBox set of UOs at the $t$-th frame, denoted by $\boldsymbol{\mathcal{X}}^{UO}_b(t)$. Then, when the new frame $t+1$ comes, the BBoxes of the same objects in two consecutive frames need to be associated, so that the UOs can be tracked in the image sequence, which is known as the MOT technique. For simplicity, we employ a simple, fast, and easy-to-deploy MOT algorithm called the ByteTrack \cite{zhang2023bytetrackv2}. Given two consecutive BBox sets, $\boldsymbol{\mathcal{X}}^{Box}_b(t)$ and $\boldsymbol{\mathcal{X}}^{Box}_b(t+1)$, the ByteTrack outputs a distinguishable BBox set $\boldsymbol{\mathcal{X}}^{DBox}_b(t+1)=\{DBox_{b,n}(t+1)|n=1,2,..., N^P_b(t+1), b \in \bm{\mathcal{B}}\}$, in which the bonding box of detected object is distinguishable with a retrievable label $id_{b,n}(t+1)$, i.e., $DBox_{b,n}(t+1)=\{ Box_{b,n}(t+1), id_{b,n}(t+1) \}$. As a result, the BBox set of UOs at the $(t+\tau)$-th frame, denoted by $\boldsymbol{\mathcal{X}}^{UO}_b(t+\tau)$, can be swiftly updated by retrieving the bounding boxes according to the labels of distinguishable BBoxes from $\boldsymbol{\mathcal{X}}^{DBox}_b(t+\tau)$, therefore avoiding frequent utilization of $\boldsymbol{\mathcal{X}}^{Bsm}_b(t+\tau)$ to identify and track UOs.
\begin{algorithm}[!tbp] 
	\caption{The two-step matching based user identification}	\label{alg1}
	\begin{algorithmic}[1]
        \STATE \textbf{Input}: \\
        \quad The BSM set $\boldsymbol{\mathcal{X}}^{Box}_b(t)$, the BBox set $\boldsymbol{\mathcal{X}}^{Bsm}_b(t)$ and the preset threshold $\gamma_1, \gamma_2 > 0$;
		\STATE  Set $\boldsymbol{\mathcal{X}}^{UO}_b(t)=\emptyset$;
        \STATE  \textbf{repeat}
        \STATE  \quad Set $\boldsymbol{\hat{\mathcal{X}}}^{Box}_b(t)=\boldsymbol{\mathcal{X}}^{Box}_b(t)$
        \STATE  \quad Randomly select a $Bsm_{b,u}(t)$ from $\boldsymbol{\mathcal{X}}^{Bsm}_b(t)$;
        \STATE  \quad \textbf{repeat}
		\STATE  \quad \quad Randomly select a $Box_{b,n}(t)$ from $\boldsymbol{\hat{\mathcal{X}}}^{Box}_b(t)$;
        \STATE  \quad \quad Obtain $y_{b,u}^P$, $z_{b,u}^P$, $d^C_{b,u}$, $h_{u}^W$ from $Bsm_{b,u}(t)$;
        \STATE  \quad \quad Obtain $u_{b,n}^P$, $v_{b,n}^P$, $q_{b,n}^P$ from $Box_{b,n}(t)$;
        \STATE  \textbf{\scriptsize Step1:} \textbf{if} $\left|u_{b,n}^P - y_{b,u}^P \right| + \left|v_{b,n}^P - z_{b,u}^P \right| < \gamma_1$ \textbf{and} \\
         \textbf{\scriptsize Step2:}  $\left|\frac{h^W_u \times f}{q^{P}_{b,n}} - d^C_{b,u}\right| < \gamma_2$ \textbf{then}
        \STATE  \quad \quad \quad $\boldsymbol{\mathcal{X}}^{UO}_b(t) \cup {Box_{b,n}(t)}$;
        \STATE  \quad \quad \textbf{end if}
        \STATE  \quad \quad $\boldsymbol{\hat{\mathcal{X}}}^{Box}_b(t)=\boldsymbol{\hat{\mathcal{X}}}^{Box}_b(t) \setminus {Box_{b,n}(t)}$;
		\STATE  \quad \textbf{until} $\boldsymbol{\hat{\mathcal{X}}}^{Box}_b(t) = \emptyset$;
        \STATE  \quad $\boldsymbol{\mathcal{X}}^{Bsm}_b(t)=\boldsymbol{\mathcal{X}}^{Bsm}_b(t) \setminus {Bsm_{b,u}(t)}$;
		\STATE  \textbf{until} $\boldsymbol{\mathcal{X}}^{Bsm}_b(t) = \emptyset$;
		\STATE  \textbf{Output} The bounding box set of user vehicles $\boldsymbol{\mathcal{X}}^{UO}_b(t)$.
	\end{algorithmic}
\end{algorithm}
\vspace{-0.5cm}
\subsection{Binary Encoding Map based Gain and Beam Prediction Network}
\begin{figure*}[tp] 
    \centering
    \includegraphics[scale=0.6]{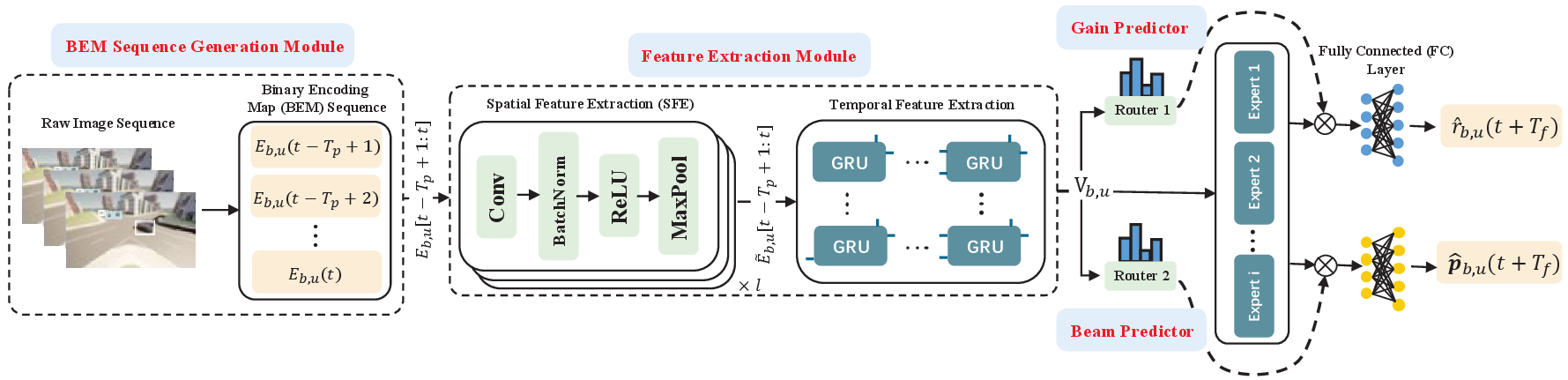}
    \caption{The diagram of the binary encoding map based gain and beam prediction network (BEM-GBPN) at the BS. }
    \label{BEM-GBPN_diagram}
    \vspace{-0.5cm}
\end{figure*}

After user identification and tracking via the OMS-UIT algorithm, the spatial and temporal information of MUs and scatterers in the physical environment can be obtained. To predict beamforming gains and optimal beams for MUs, the BEM-GBPN is proposed and deployed at each BS, as shown in Fig. \ref{BEM-GBPN_diagram}, which contains four modules: the BEM sequence generation module, the feature extraction module, the beamforming gain predictor, and the optimal beam predictor. 
\subsubsection{BEM Sequence Generation Module}
\label{Img_pre}
\begin{figure}[!tbp] 
    \centering
    \includegraphics[scale=0.5]{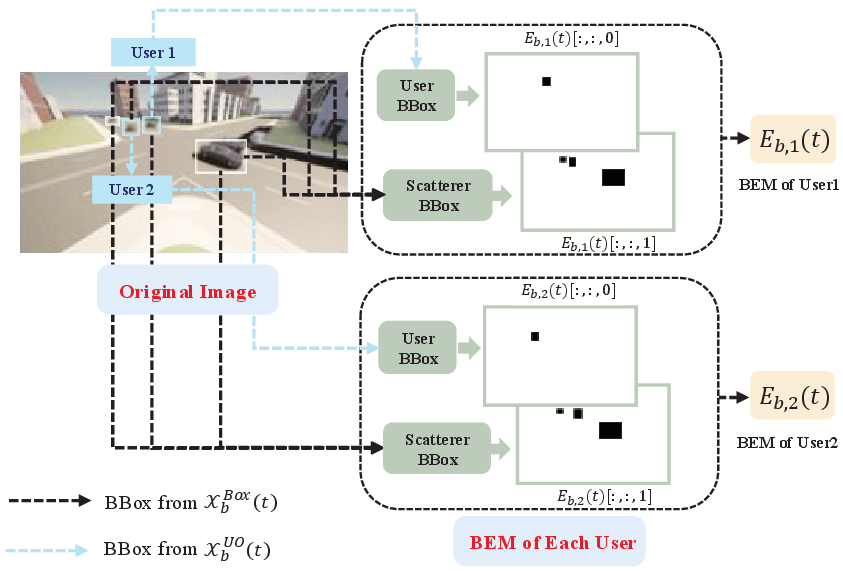}
    \caption{The diagram of the binary encoding map (BEM) construction. }
    \label{BEM_img}
    \vspace{-0.5cm}
\end{figure}

Since the background characteristics (i.e., the buildings, roads, trees, etc.) remain stationary for a fixed camera, their impact on the UO's channel follows an almost fixed distribution. In the context of vehicular networks, the primary influence on the UO's channel comes from other moving targets (e.g., vehicles). To effectively represent the spatial characteristics of the environment and dynamic objects, we present the BEM-based representation method. Specifically, as shown in Fig. \ref{BEM_img}, the BEM for the $t$-th frame, denoted by $\boldsymbol{E}_{b,u}(t) \in \mathbb{R}^{W \times H \times 2}$, is generated by assigning a value of $1$ to regions covered by BBoxes in $\boldsymbol{\mathcal{X}}^{Box}_b(t)$ corresponding to all dynamic objects (i.e., PUOs), while encoding uncovered background as $0$. Moreover, the UO's BBox is encoded into the first channel of the BEM, i.e., $\boldsymbol{E}_{b,u}(t)[:,:,0]$, while scatterers' BBoxes are encoded into the second channel of the BEM, i.e., $\boldsymbol{E}_{b,u}(t)[:,:,1]$. Finally, we can construct the BEM sequence $\boldsymbol{E}_{b,u}[t-T_p+1:t]=[\boldsymbol{E}_{b,u}(t-T_p+1), \boldsymbol{E}_{b,u}(t-T_p+2), ..., \boldsymbol{E}_{b,u}(t)]$ for $T_p$ consecutive frames.
\subsubsection{Feature Extraction Module}
\label{feature_extraction}
Subsequently, a feature extraction module is proposed to extract both spatial and temporal features of the BEM sequence $\boldsymbol{E}_{b,u}[t-T_p+1:t]$. For the spatial feature, it is worth noting that, the size of bounding boxes in the BEM reflects the relative distance between objects and the camera, while the relative positions of bounding boxes represent the spatial distribution of objects. Since a convolutional neural network (CNN) \cite{ConvNN} with appropriate convolutional kernel sizes and strides has a sufficiently large perception field to extract the features of objects' distance and relative position, we propose a CNN-based spatial feature extraction (SFE) module to extract and compress these spatial features. Specifically, a CNN cell is constructed to extract features that contain semantic information to distinguish users and scatterers in terms of relative positions and distances via multi-channel feature extraction. The CNN cell, which is able to conduct multi-channel feature extraction, can be mathematically expressed as
\begin{equation}\label{eq9}
\text{CNNCell}(\mathbf{x})=\text{MaxPool}(\text{ReLU}(\text{BatchNorm}(\text{Conv}(\mathbf{x})))),
\end{equation}
\noindent where $\text{Conv}(\cdot)$ represents the convolutional layer, $\text{BatchNorm}(\cdot)$ denotes the batch normalization layer that normalizes the inputs of the layers by re-centering and re-scaling, $\text{ReLU}(\cdot)$ is the Rectified Linear Unit (ReLU) nonlinear activation function, and $\text{MaxPool}(\cdot)$ is the max pooling function to down-sample the input representation. 

Subsequently, the embedded BEM sequence of the $u$-th user can be extracted by $l$ CNN cells as
\begin{multline}\label{eq10}
\boldsymbol{\tilde{E}}_{b,u}[t-T_p+1:t]= \\
\text{FC}\left(\text{CNNCell}^{l}\left(\cdots \text{CNNCell}^{1}\left(\boldsymbol{E}_{b,u}[t-T_p+1:t]\right)\right)\right),
\end{multline} 
where the $\text{FC}(\cdot)$ represents the fully connected (FC) layer utilized to map the input feature vector to a new feature, which can be formulated as
\begin{equation}\label{eq11}
\text{FC}(\mathbf{x})=\text{ReLU}(\mathbf{W}_f \cdot \mathbf{x}+\textbf{b}_f),
\end{equation}
where $\mathbf{x}$, $\mathbf{W}_f$, and $\textbf{b}_f$ are input feature, learnable weight matrix and bias vector of the FC layer, respectively. Since CNNs inherently fail to capture sequential dependencies in input data, they are not expected to effectively learn the temporal features within the input embedded BEM sequence. To overcome this, a temporal feature extractor consisting of $T_p$ gated recurrent units (GRUs) is proposed to learn the temporal features within the embedded BEM sequence $\boldsymbol{\tilde{E}}_{b,u}[t-T_p+1:t]$. Denoting the output, and hidden state of the $\tau$-th GRU unit as $\mathbf{O}(\tau)$, and $\mathbf{H}(\tau)$, respectively, a reset gate $\boldsymbol{R}(\tau)$ is used to discard some of the historical hidden states, and an update gate $\boldsymbol{Z}(\tau)$ controls the ratio of new input and historical information being used for the output, which can be formulated as 
\begin{equation}\label{eq12}
    \boldsymbol{R}(\tau)=\sigma(\mathbf{W}_{xr} \boldsymbol{\tilde{E}}_{b,u}(\tau) + \mathbf{W}_{hr}\mathbf{H}(\tau-1)+\textbf{b}_r),
\end{equation}
\begin{equation}\label{eq13}
    \boldsymbol{Z}(\tau)=\sigma(\mathbf{W}_{xz} \boldsymbol{\tilde{E}}_{b,u}(\tau) + \mathbf{W}_{hz}\mathbf{H}(\tau-1)+\textbf{b}_z),
\end{equation}
where $\tau \in [t-T_p+1, t]$, and $\sigma(\cdot)$ is the sigmoid function, i.e.,
\begin{equation}\label{eq14}
    \sigma(x) = \frac{1}{1+e^{-x}} \in (0,1),
\end{equation}
and $(\mathbf{W}_{xr}$, $\mathbf{W}_{hr})$ and $(\mathbf{W}_{xz}$, $\mathbf{W}_{hz})$ are the learnable weight matrices of the reset gate and update gate, and $\textbf{b}_{r}$ and $\textbf{b}_{z}$ are the corresponding bias vectors, respectively. 

Subsequently, the candidate hidden state, which is an intermediate variable generated by combining the current input and gated historical information, provides a candidate value for updating the final hidden state. It is formulated as 
\begin{equation}\label{eq15}
    \tilde{\mathbf{H}}(\tau) = tanh(\mathbf{W}_{xh} \boldsymbol{\tilde{E}}_{b,u}(\tau) + \mathbf{W}_{hh}(\boldsymbol{R}(\tau) \odot \mathbf{H}(\tau-1)) + \mathbf{b}_{h}),
\end{equation}
where $tanh(\cdot)$ is the hyperbolic tangent activation function, i.e.,
\begin{equation}
    tanh(x)=\frac{e^x-e^{-x}}{e^x+e^{-x}} \in (-1,1).
\end{equation}

The output of the $\tau$-th GRU unit is a weighted combination of historical and new information determined by the update gate as
\begin{equation}\label{eq16}
    \mathbf{O}(\tau) = \mathbf{Z}(\tau) \odot \mathbf{H}(\tau-1) + (1-\mathbf{Z}(\tau)) \odot \tilde{\mathbf{H}}(\tau).
\end{equation}
Finally, the spatio-temporal features at the $t$-th frame extracted by the feature extraction module can be expressed as $\boldsymbol{V}_{b,u}=\mathbf{O}(t)$.

Then, $I$ feature mappers, referred as experts and denoted by $\text{EXP}_i, i=1,..., I$, are designed to project $\boldsymbol{V}_{b,u}$ into different kinds of semantic features, including the distance, relative position, and trajectory of the $u$-th user and surrounding scatterers. The semantic feature extracted by the $i$-th expert can be represented as
\begin{equation}\label{eq18}
\boldsymbol{K}_i=\text{EXP}_{i}\left(\boldsymbol{V}_{b,u}\right).
\end{equation}
These semantic features can be utilized for the BS to perceive whether the user is approaching, moving away, or possibly being obstructed by the nearby scatterers in future moments. 
\subsubsection{Beamforming Gain Predictor}
It is worth emphasizing that even when many semantic features are extracted by various experts, it is still challenging to adaptively combine these semantic features towards different tasks. For instance, the distance between the BS and the user might dominate other semantic features like user trajectory in beamforming gain prediction, while user trajectory might dominate the distance or relative position in optimal beam prediction. To address this challenge, we further propose an MoE-based beamforming gain predictor to adaptively evaluate the importance of extracted semantic features and combine them for beamforming gain prediction. Specifically, the MoE structure is mainly composed of a weighting router generating the importance of all semantic features as
\begin{equation}\label{eq19}
\boldsymbol{\eta}_1= \boldsymbol{W}_{router1} \cdot \boldsymbol{V}_{b,u},
\end{equation}
where $\boldsymbol{W}_{router1}$ is the learnable mapping matrix of the router, $\eta_1=[\eta_{1,1}, \eta_{1,2}, ..., \eta_{1,I}]$ is the weighting vector, with $\eta_{1,i}$ representing the importance of the semantic feature $\boldsymbol{K}_i$ for beamforming gain prediction. 

Note that in practical scenarios, the beamforming gain of a user tends to fluctuate within a defined range, whose upper and lower bounds can be roughly estimated according to the measured values in practice, denoted as $P_{max}$ and $P_{min}$, respectively. To enhance the predictive capacity of the beamforming gain predictor, the beamforming gain is normalized within an interval $[0,1]$ as the training label. In essence, the normalization procedure can be represented as
\begin{equation} \label{eq20} 
\begin{split}
    r'_{b,u}(t)=\frac{r_{b,u}(t)-P_{min}}{P_{max}-P_{min}},
\end{split}
\end{equation}
where
\begin{equation} \label{eq21}
\begin{split}
    r_{b,u}(t)={\left|(\boldsymbol{h}_{b,u}(t))^H \boldsymbol{F}^{RF*}_{b}(t) [\boldsymbol{F}^{BB*}_{b}(t)]_{:,u}\right|^2},
\end{split}
\end{equation}
where $\boldsymbol{F}^{RF*}_{b}(t)$ and $\boldsymbol{F}^{BB*}_{b}(t)$ are the optimal analog beamformer and digital beamformer, respectively, $r_{b,u}(t)$ represents the optimal beamforming gain, and $\boldsymbol{h}_{b,u}(t)$ is the channel between the $b$-th BS and the $u$-th user at the $t$-th frame. 

Subsequently, a prediction layer is utilized to predict beamforming gains at the $(t+T_f)$-th frame given the semantic features and weighting vector $\boldsymbol{\eta}_1$ as 
\begin{equation}\label{eq22}
\hat{r}'_{b,u}(t+T_f) = \sigma\left(\text{FC}\left(\sum_{i \in I} \eta_{1,i} 
\cdot \boldsymbol{K}_{i}\right)\right), \eta_{1,i} \in \boldsymbol{\eta}_1.
\end{equation}

The training process is conducted offline, and the Mean Squared Error (MSE) is selected as the loss function, which can be expressed as
\begin{equation}\label{eq23}
Loss_m = (r'_{b,u}(t+T_f) - \hat{r}'_{b,u}(t+T_f))^2.
\end{equation}
During evaluation, the estimated beamforming gain $\hat{r}'_{b,u}(t+T_f)$ is restored to the original range using $P_{max}$ and $P_{min}$ as $\hat{r}_{b,u}(t+T_f)$.
\begin{figure}[!tbp]
    \centering
    \includegraphics[scale=0.4]{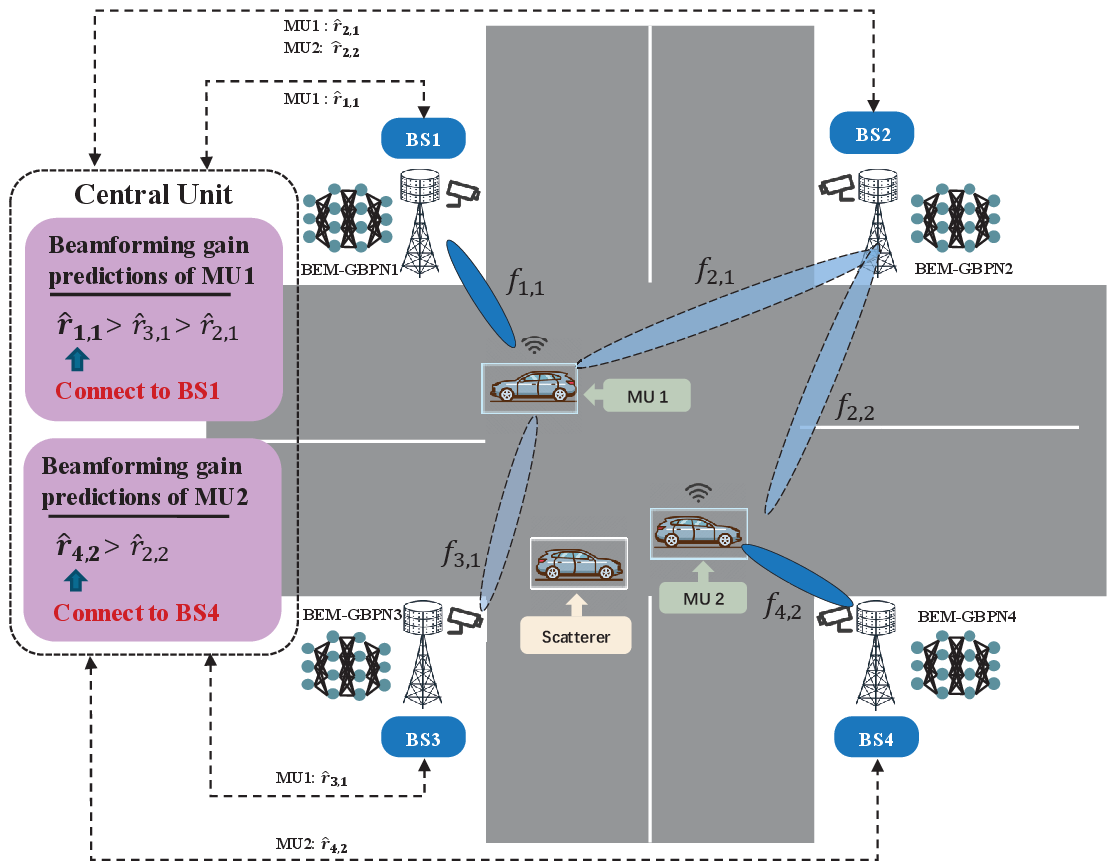}
    \caption{The illustration of the proposed proactive BS selection and beam switching approach.}
    \label{BS_handoff_framework}
    \vspace{-0.5cm}
\end{figure}
\subsubsection{Optimal Beam Predictor}
Similar to the beamforming gain predictor, we propose a MoE-based optimal beam predictor for accurate optimal beam prediction, where another weighting router $\boldsymbol{W}_{router2}$ is utilized to evaluate the importance of extracted semantic features for beam prediction, and generate the weighting vector for beam prediction, i.e., $\eta_2=[\eta_{2,1}, \eta_{2,2}, ..., \eta_{2,I}]$, similar to (\ref{eq19}). Then, another prediction layer is exploited to predict the optimal beams at the $(t+T_f)$-th frame as
\begin{equation}\label{eq24}
\hat{\boldsymbol{p}}_{b,u}(t+T_f) = \sigma\left(\text{FC}\left(\sum_{i \in I} \eta_{2,i} 
\cdot \boldsymbol{K}_{i}\right)\right), \eta_{2,i} \in \boldsymbol{\eta}_2,
\end{equation}
where $\hat{\boldsymbol{p}}_{b,u}(t+T_f)=[ \hat{p}_1, \hat{p}_2,..., \hat{p}_{N_{T}} ]$ is the probability vector, where $\hat{p}_n$ denotes the predicted probability of the $n$-th beam being the optimal beam at the $(t+T_f)$ frame. The loss function for training the optimal beam predictor is the binary cross-entropy defined as
\begin{subequations}\label{eq25}
    \begin{align}
        & Loss_c =  - \sum\limits_{n = 1}^{{N_{T}}} {{p_n}\log(\hat{p}_n)}, \\
        & \hat{p}_n \in \hat{\boldsymbol{p}}_{b,u}(t+T_f), p_n \in \boldsymbol{p}_{b,u}(t+T_f), 
    \end{align}
\end{subequations} 
where $\boldsymbol{p}_{b,u}(t+T_f)=[p_1, p_2,..., p_{N_{T}}] \in \{0,1\}^{N_{T}}$ is the one-hot label of the optimal beam. Specifically, if the $n$-th beam in $\bm{\mathcal{F}}$ is the optimal beam, then $p_n = 1$ otherwise $p_n = 0$. Given the predicted optimal beam, the analog beamformer $\boldsymbol{F}^{RF}_{b_u}$ in (\ref{eq4}) can be directly obtained without intensive beam sweeping. 
\subsection{Proactive BS Selection and Beam Switching Approach}
In mmWave wireless networks, LOS link blockage and severe path loss have long been acknowledged as critical challenges. Therefore, implementing effective intra-BS beam switching and proactive inter-BS handoff is essential to improve the total transmission rate \cite{giordani2016multiconnectivity}. 

Since the channel coherence time is short, especially in vehicular networks with high mobility, leveraging pilots to measure the link qualities of MUs and then performing mobility management leads to unaffordable pilot overhead and delay. Therefore, a proactive BS selection and beam switching approach based on the predictions from BEM-GBPN is proposed to mitigate the pilot overhead and delay, which is shown in Fig. \ref{BS_handoff_framework}. Specifically, the BSs first utilize BEM-GBPN to independently predict the beamforming gains and optimal beams for MUs within the view. For instance, BS1 is able to perceive MU1, while BS2 is able to perceive MU1 and MU2. Then the BSs transmit the prediction results to the central unit via the optical/wireless backhaul links, which contain only a minimal amount of data, i.e., the identity of detected MUs and the corresponding predicted beamforming gains. The central unit compares the predicted beamforming gains of an MU for all BSs and controls the BS that is most likely to offer the highest beamforming gain to serve this MU. Finally, the serving BS configures its analog beamformer based on the beam prediction for the MU. The process of BS selection can be expressed as 
\begin{equation}\label{eq26}
     b_u = \underset{b \in \{1,2,...,B \} } {\arg\max} \hat{r}_{b,u}(t+T_f),
\end{equation}
and the process of beam switching can be formulated as 
\begin{subequations}\label{eq27}
\begin{align}
     &[\boldsymbol{F}^{RF}_{b_u}]_{:,u}(t+T_f) = \boldsymbol{f}_n \in \bm{\mathcal{F}}, \\
     &n = \underset{i \in \{1,2,...,N_{T} \} } {\arg\max} \hat{p}_i \in \hat{\boldsymbol{p}}_{b_u,u}(t+T_f).
\end{align}
\end{subequations}
In addition, to eliminate multiuser interference \cite{sun2019beam}, the digital beamformer can be designed under the power constraint (\ref{eq4d}), i.e., $\boldsymbol{F}^{BB}_{b_u}=[\boldsymbol{f}^{BB}_1, \boldsymbol{f}^{BB}_2, ..., \boldsymbol{f}^{BB}_{U}]$, where $\boldsymbol{f}^{BB}_u \in \mathbb{R}^{N_{RF} \times 1}$ is a vector with binary elements, indicating that if the transmitted signal to the $u$-th user is matched to the $i$-th RF chain, the corresponding element $[\boldsymbol{F}^{BB}_{b_u}]_{i,u}$ is set to 1, otherwise 0. 
\begin{figure} [tp!] 
 \centering
 \subfloat[\label{sim_env}The 3D model of the scenario in CARLA.]{\hspace{2mm}
  \includegraphics[width=0.45\linewidth]{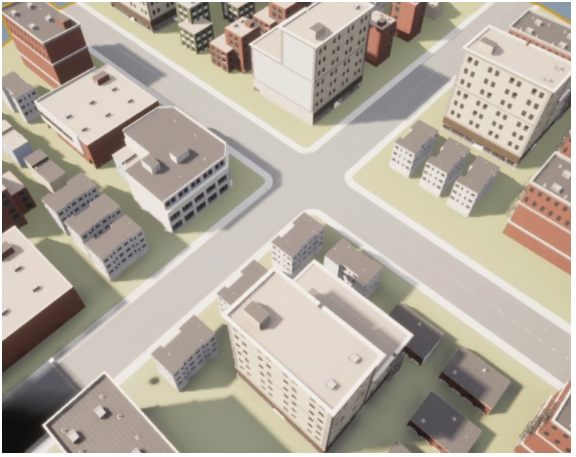}} 
 \subfloat[\label{WI_env}The 3D model of the scenario in Wireless Insite.]{\hspace{2mm}
  \includegraphics[width=0.45\linewidth]{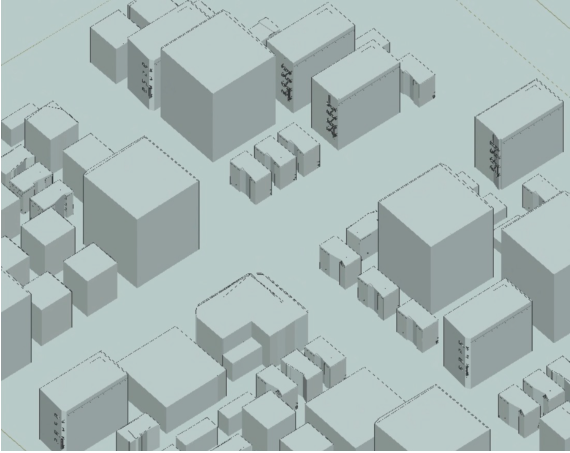}} 
 \caption{The simulated environment in CARLA and Wireless Insite. The environmental buildings and vehicles in Wireless Insite are synchronized with those in CARLA.}
\end{figure}
\section{Simulation Results}
\subsection{Simulation Setup}
To evaluate the performance of the proposed OMS-based mobility management scheme, we simulate the communication environment with moving vehicles (some of which are MUs) to collect OOB modality data and wireless channel data simultaneously. The detailed simulation setups are illustrated as follows. As shown in Fig. \ref{sim_env}, a V2I communications scenario at a crossroad, with $B=4$ BSs positioned at each corner serving vehicular users, is simulated with the CARLA simulation software \cite{CARLA_software}. The coordinates of the four vertices of the considered crossroad are set to [(-175,25), (-125,25), (-125,-25), (-175,-25)] in CARLA. Then, we adopt random types of vehicles with different sizes and appearances, including cars, minivans, buses, trucks, etc., and random driving speeds with a maximum limitation of $40$km/h to verify the generalization capability of our proposed mobility management scheme. 
\subsubsection{OOB Modality Data Generation}
The OOB modality data, including visual sensing data and location data, are captured at every simulation time frame, where the sampling interval between two frames is $T_s=20$ms. Specifically, an RGB camera is installed on the top of each BS and facing toward the crossroad, and all the cameras installed on the $B$ BSs take images simultaneously. The parameters of cameras are shown in Table \ref{Table1}, and the image is resized from a resolution of $1280\times720$ to $480\times320$ to reduce the computational overhead. We collect an image set $\bm{\mathcal{I}}_t$, which contains RGB images taken by these $B$ cameras at the $t$-th frame. In addition, all vehicles’ parameters, including their types, speeds, and locations, are automatically controlled by the SUMO software \cite{SUMO_software} to generate random traffic flow. Therefore, we can collect a BSM set $\bm{\mathcal{B}}_t$ by directly extracting identities, locations, and sizes of all vehicles from the SUMO, in which $Bsm_{n}(t) \in \bm{\mathcal{B}}_t$ represents the BSM of the $n$-th vehicle at the $t$-th frame.
\begin{table*}[ht]
\centering
\caption{Simulation Parameters}
\label{Table1}
\resizebox{1\textwidth}{!}{
\begin{tabular}{c|c|c|c|c|c|c|c|c|c}
\hline
\multicolumn{10}{c}{Parameters of cameras in CARLA} \\ \hline
Camera & Loc-x/m & Loc-y/m & Loc-z/m & Pitch/\textdegree & Yaw/\textdegree & Roll/\textdegree & Image Width/pixel & Image Height/pixel & Fov/\textdegree \\ \hline
Camera 1 & -162 & 12 & 4 & -25 & 45 & 0 & 1280 & 720 & 120 \\ \hline
Camera 2 & -162 & -12 & 4 & -25 & -45 & 0 & 1280 & 720 & 120 \\ \hline
Camera 3 & -138 & 12 & 4 & -25 & 135 & 0 & 1280 & 720 & 120 \\ \hline
Camera 4 & -138 & -12 & 4 & -25 & -135 & 0 & 1280 & 720 & 120 \\ \hline
\multicolumn{10}{c}{Parameters of Wireless Insite for ray tracing} \\ \hline
\multicolumn{2}{c|}{Paths} & 5 & \multicolumn{4}{|c|}{Reflections} & 6 & Diffractions & 1 \\ \hline
\multicolumn{2}{c|}{Propagation Model} & X3D & \multicolumn{4}{|c|}{Building Material} & Concrete & Vehicle Material & Metal \\ \hline
\end{tabular}
}
\end{table*}

\subsubsection{Wireless Channel Data Generation}
\label{WD_gen}
The Wireless Insite software \cite{WI_software} is utilized to generate the attenuation/angle/delay parameters of signal paths to obtain wireless channels according to (\ref{eq2}). Since mmWave channels have a high spatial correlation with the constructed environment, environmental buildings and vehicles in Wireless Insite are completely synchronized with those in the CARLA, as shown in Fig. \ref{WI_env}. Note that the carrier frequency for downlink communications is $28$GHz. Each BS is equipped with a ULA, with $N_{T}=64$ antenna elements and $N_{RF}=4$ RF chains, and each vehicle is equipped with an omnidirectional antenna. In addition, the height of the ULA equipped at each BS is set $0.5$m higher than the corresponding camera, and the omnidirectional antenna of each vehicle is set at $0.05$m above the roof center of the vehicle. At the $t$-th frame that the $B$ cameras take images, we use Wireless Insite to generate a channel set $\bm{\mathcal{H}}_t$, containing channels between all BSs and all vehicles in the crossroad according to the simulation parameters in Table \ref{Table1}, in which $\boldsymbol{h}_{b,n}(t) \in \bm{\mathcal{H}}_t$ represents a channel between the $b$-th BS and the $n$-th vehicle at the $t$-th frame. 
\begin{figure}[!tbp]
    \centering
    \includegraphics[scale=0.4]{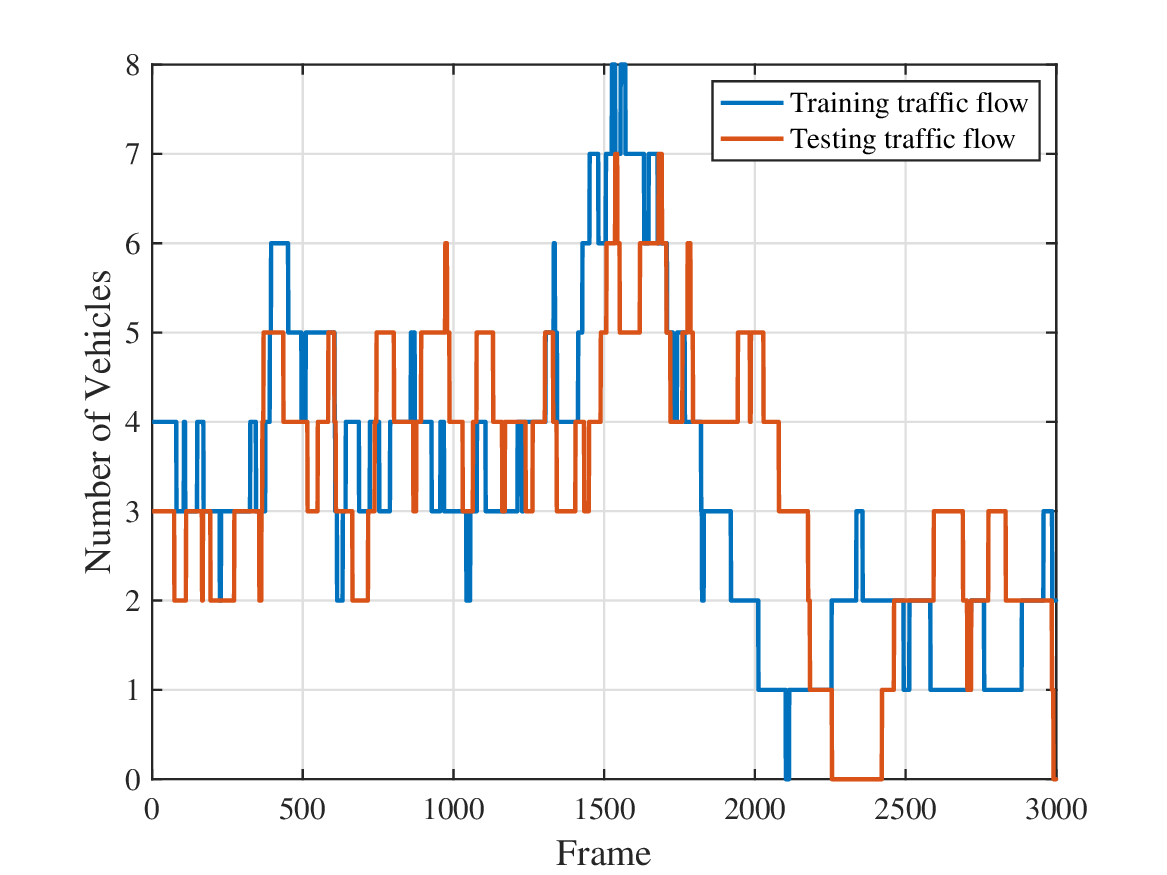}
    \caption{The distribution of the number of vehicles over frames in the training traffic flow and testing traffic flow. } 
    \label{veh_dis}
    \vspace{-0.5cm}
\end{figure}
\subsubsection{Dataset Generation}
To simulate the traffic flow, we randomly generate $30$ vehicles at the edge of traffic lanes for initialization. Data collection begins once the vehicles enter the considered crossroad. Since vehicles' trajectories and spatio-temporal distributions are affected by their initial conditions (e.g., speeds and locations), we carry out two different traffic flows with different vehicle initializations, called the training traffic flow and the testing traffic flow, to construct the training dataset $\bm{\mathcal{C}}^{t\&v}$ and the testing dataset $\bm{\mathcal{C}}^{test}$, respectively. Each traffic flow contains $3000$ frames (i.e., $60$s), as shown in Fig. \ref{veh_dis}, and the distribution of the number of vehicles over frames differs between the two traffic flows. The BEM-GBPN is trained on the training dataset and then evaluated on the testing dataset with fixed parameters, which effectively verifies the generalization capability of the proposed algorithm. 
\begin{figure}[!tbp]
    \centering
    \includegraphics[scale=0.4]{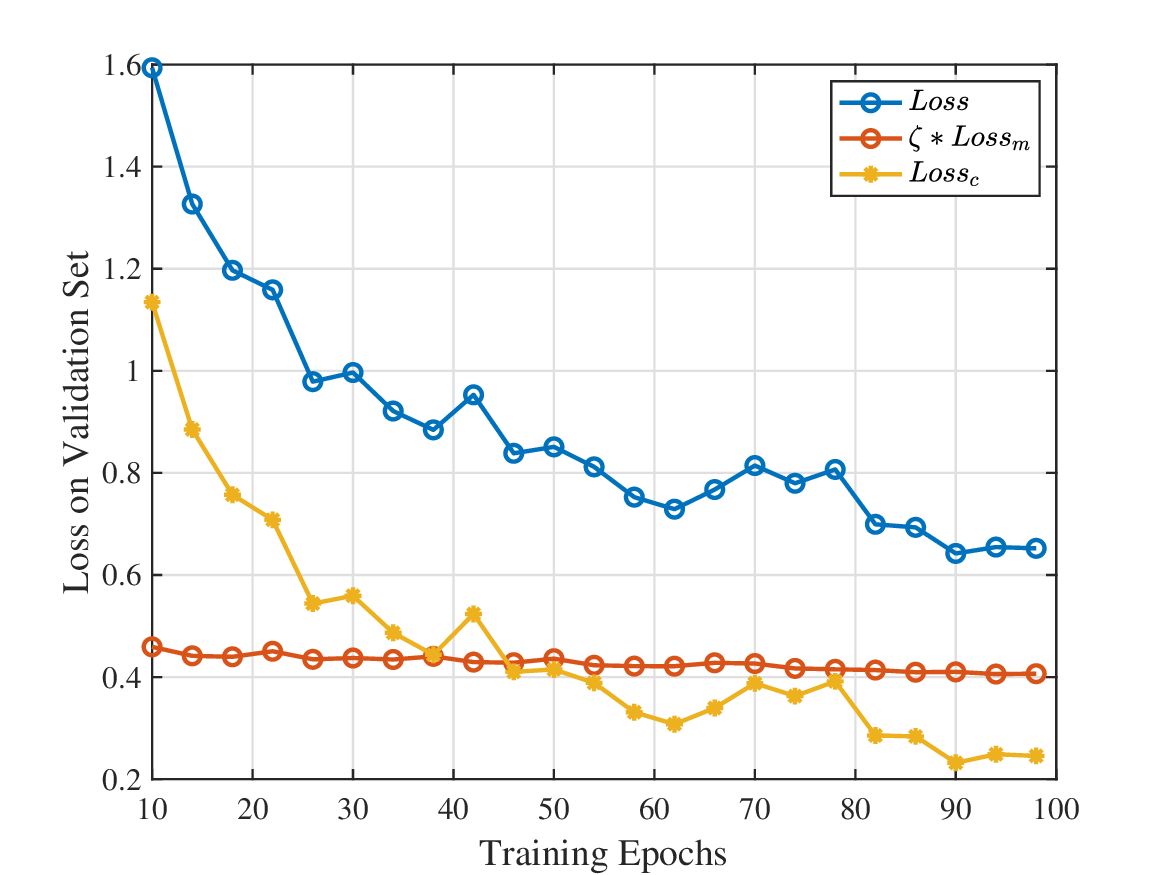}
    \caption{The validation losses of BEM-GBPN versus the number of training epochs. } 
    \label{loss_curve}
    \vspace{-0.5cm}
\end{figure}

Specifically, the training dataset $\bm{\mathcal{C}}^{t\&v}$ is generated as follows. First of all, the training traffic flow is chosen for simulation. For the $b$-th BS, $b=1,2,..., B$, each vehicle exists in the crossroad and can be visually detected by the $b$-th BS during $(T_f+T_p)$ consecutive frames is configured as a UO to collect a batch of training samples $\bm{\mathcal{C}}_{b,n}^{t\&v} \in \bm{\mathcal{C}}_{b}^{t\&v}, n=1,2,...,30$. Each sample in the $\bm{\mathcal{C}}_{b,n}^{t\&v}$ contains the BEM sequence $\boldsymbol{E}_{b,n}[t-T_p+1:t]$ as the input data, and the beamforming gain $r_{b,n}(t+T_f)$ and the one-hot label of the optimal beam $\boldsymbol{p}_{b,n}(t+T_f)$ as labels to train the BEM-GBPN. The BEM sequence can be generated from the images set sequence $\bm{\mathcal{I}}_{t-T_p+1}, \bm{\mathcal{I}}_{t-T_p+2}, ..., \bm{\mathcal{I}}_t$. The labels of the beamforming gain and optimal beam can be generated from the channel sequence $\bm{\mathcal{H}}_{t+1}, \bm{\mathcal{H}}_{t+2}, ..., \bm{\mathcal{H}}_{t+T_f}$ via beam searching. Note that the training process is offline, and there is no need to perform beam searching in the inference phase. Moreover, $\bm{\mathcal{C}}_{b}^{t\&v}$ is randomly divided into two subsets $\bm{\mathcal{C}}_{b}^{train}$ and $\bm{\mathcal{C}}_{b}^{valid}$ to train and validate the BEM-GBPN.

Furthermore, in order to evaluate the performance of multi-user beam prediction and proactive BS selection, the testing traffic flow is chosen to generate the testing dataset $\bm{\mathcal{C}}^{test}$. We denote the number of vehicles in the crossroad at the $t$-th frame as $V_t$, which is also referred to PUOs, and the number of UOs is configured as $U$. When the $t'$-th frame meets the requirement that $V_{t'} \ge U$, $U$ UOs are selected from these $V_{t'}$ vehicles, and $\frac{V_{t'}!}{(V_{t'}-U)!U!}$ different combinations of UOs can be obtained. For each combination, the BEM sequence, the beamforming gain and optimal beam labels, corresponding to $U$ UOs and $B$ BSs, can be collected synchronously to construct a batch of testing dataset $\bm{\mathcal{C}}_{U, M,c}^{test} \in \bm{\mathcal{C}}^{test}$, $c=1,2,...,\frac{V_{t'}!}{(V_{t'}-U)!U!}$, and $M$ represents the number of interval frames between two user identification processes utilized to evaluate the performance of user tracking. 
\begin{figure}[t]
    \centering
    \includegraphics[scale=0.4]{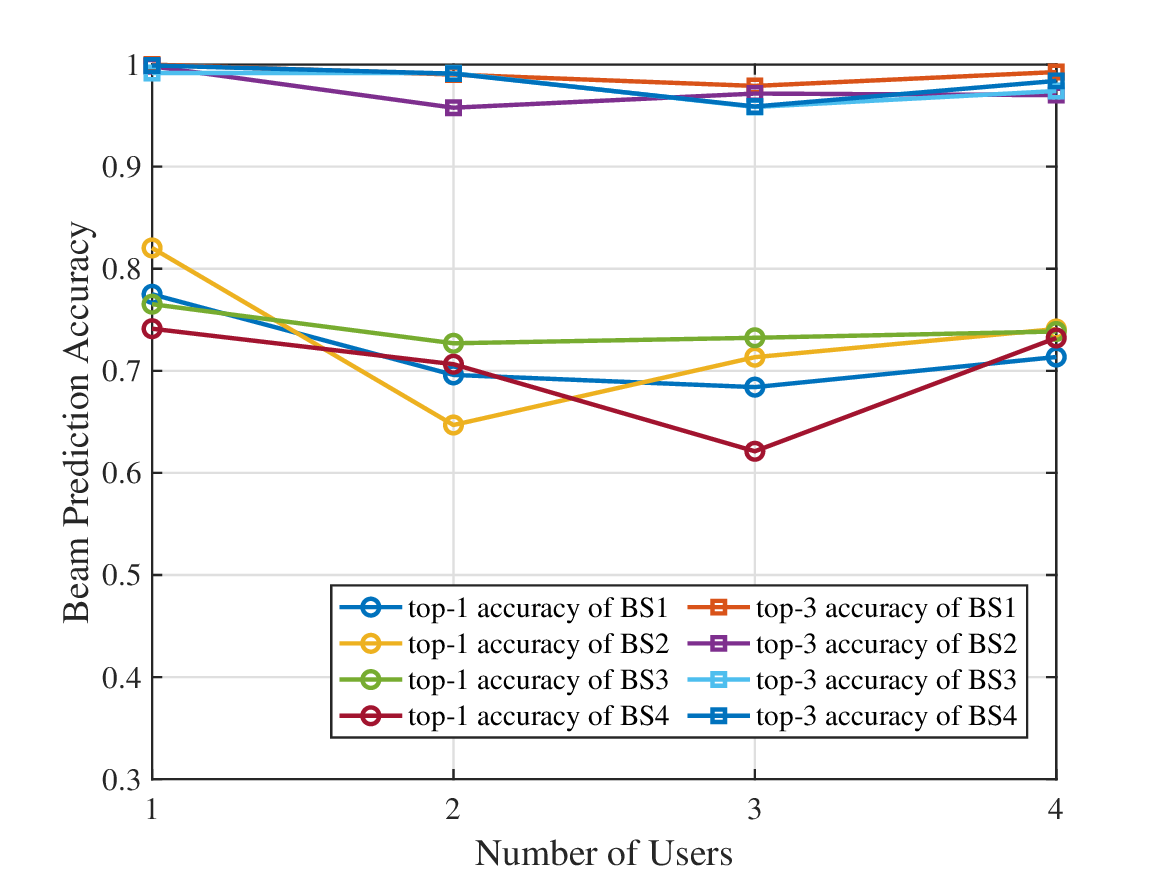}
    \caption{The beam prediction accuracy of the BEM-GBPN for all 4 BSs under different numbers of users. } 
    \label{acc_nums_ue}
\end{figure}
\begin{figure}[t]
    \centering
    \includegraphics[scale=0.4]{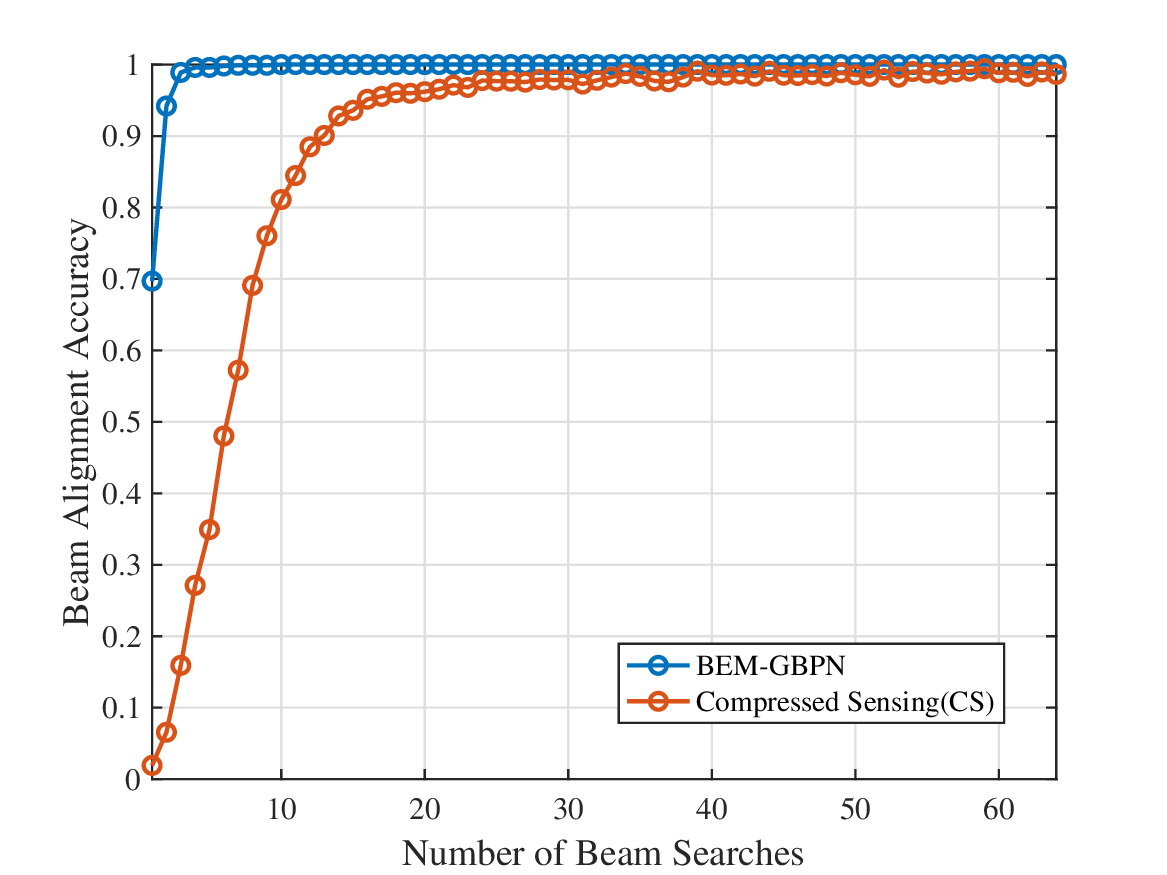}
    \caption{The beam alignment accuracy of the BEM-GBPN and CS-based beam training for BS1 under different numbers of beam searches. } 
    \label{acc_nums_searches}
    \vspace{-0.5cm}
\end{figure}
\begin{figure}[t]
    \centering
    \includegraphics[scale=0.4]{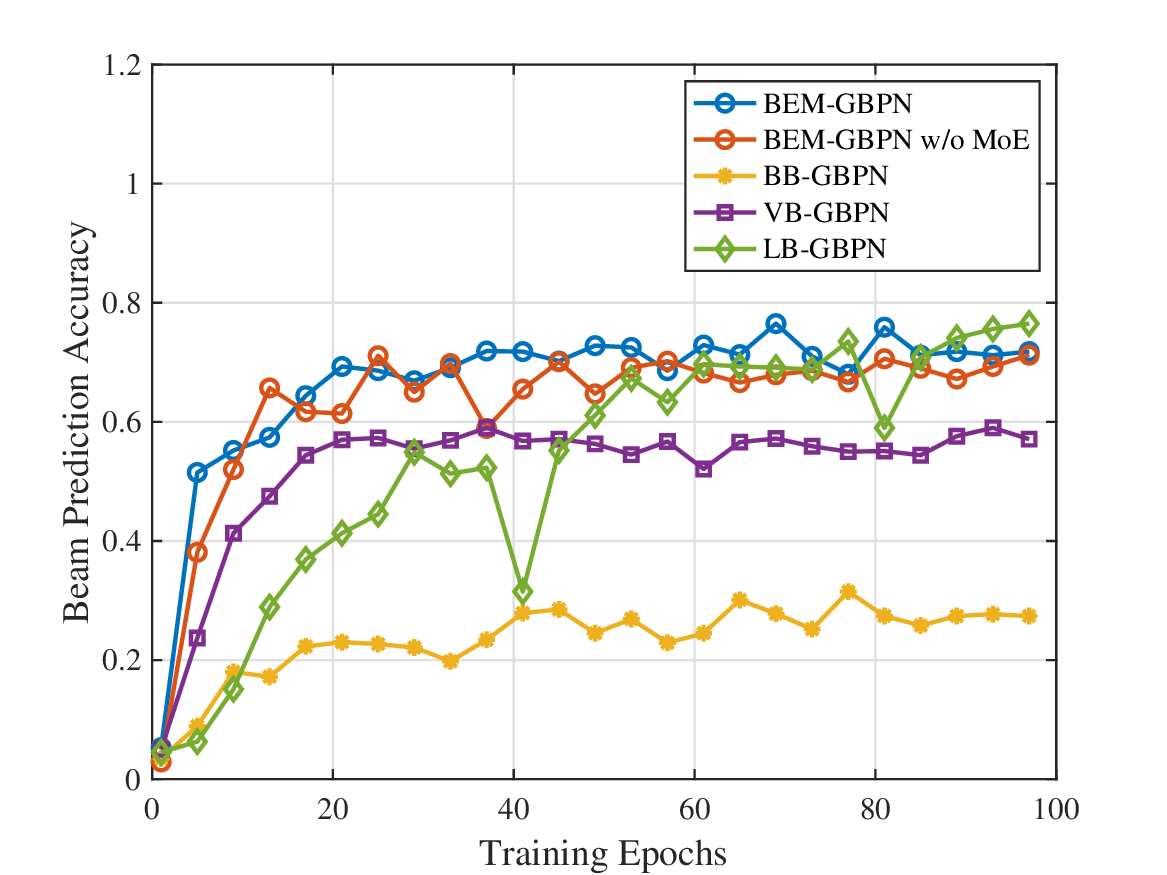}
    \caption{The beam prediction accuracy of the BEM-GBPN, BB-GBPN, VB-GBPN, and LB-GBPN for BS1 versus training epochs on the testing dataset.}
    \label{acc_epochs}
\end{figure}
\subsection{Hyper-Parameter Settings}
The hyper-parameters for the proposed OMS-based mobility management scheme are set as follows: The sequence length for temporal feature extraction is set as $T_p=20$, i.e., $400ms$, and the predicting future frame $T_f=10$, i.e., $200ms$. The size of training dataset $\bm{\mathcal{C}}_{b}^{t\&v}$ for the $b$-th BS is $5000$ samples, with $4500$ samples in $\bm{\mathcal{C}}_{b}^{train}$ for training and $500$ samples in $\bm{\mathcal{C}}_{b}^{valid}$ for validation. The size of every testing subset $\bm{\mathcal{C}}_{U,M}^{test}$ is $500$, where $U=1,2,3,4$, the BSM feedback interval $M=1,10,20,...,100$. 

The threshold $\gamma_1$ in the two-step matching based user identification is set as the summation of the width and the height of the $n$-th detected bounding box, i.e., $\gamma_1=e^P_{b,n}+q^P_{b,n}$, where $(e^P_{b,n},q^P_{b,n}) \in Box_{b,n}$, and the threshold $\gamma_2=3$. The size of BEM is set as $W=480$ and $H=320$. The SFE in the BEM-GBPN is configured with $4$ CNN cells to extract the spatial feature of the BEM. The (in channels, out channels, kernel size, stride) parameters of these four CNN cells are $(2, 4, 11, 2)$, $(4, 16, 7, 2)$, $(16, 64, 5, 1)$, and $(64, 256, 3, 1)$, respectively. The size of maxpooling is $(2 \times 2)$. The FC layer after the CNN cells has $128$ neurons. The input size of the GRU for the embedded BEM sequence is $128$. The hidden layer’s size and the number of hidden layers are set to $256$ and $2$, respectively. The number of experts in the MoE structure is $3$. Each expert is a two-layer neural network: a hidden layer with $256$ neurons followed by an output layer with $128$ neurons. Subsequently, we train the BEM-GBPN for $100$ epochs with a learning rate of $0.001$, batch size of $128$, and Adam as the optimizer. The loss function can be expressed as
\begin{equation}\label{loss_func}
Loss = \zeta * Loss_m + Loss_c, 
\end{equation}
where $Loss_m$ is the MSE loss for beamforming gain prediction and $Loss_c$ is the inary cross-entropy loss for optimal beam prediction. $\zeta=100$ is utilized to balance the orders of magnitude between the losses of the two tasks. As shown in Fig. \ref{loss_curve}, as the number of training epochs increases, the validation losses of both tasks decrease steadily and eventually converge. After training, the BEM-GBPN with the minimal validation loss is selected for testing.
\subsection{Performance Evaluation}
Since the OMS-UIT algorithm serves as the front-end of the OMS-based mobility management scheme, the accuracy of user identification and tracking can be reflected in the performance of multi-user beam prediction and BS selection. Thus, we only evaluate the accuracy of beam prediction and the performance of BS handoff and beam switching. 
\begin{figure}[t]
    \centering
    \includegraphics[scale=0.4]{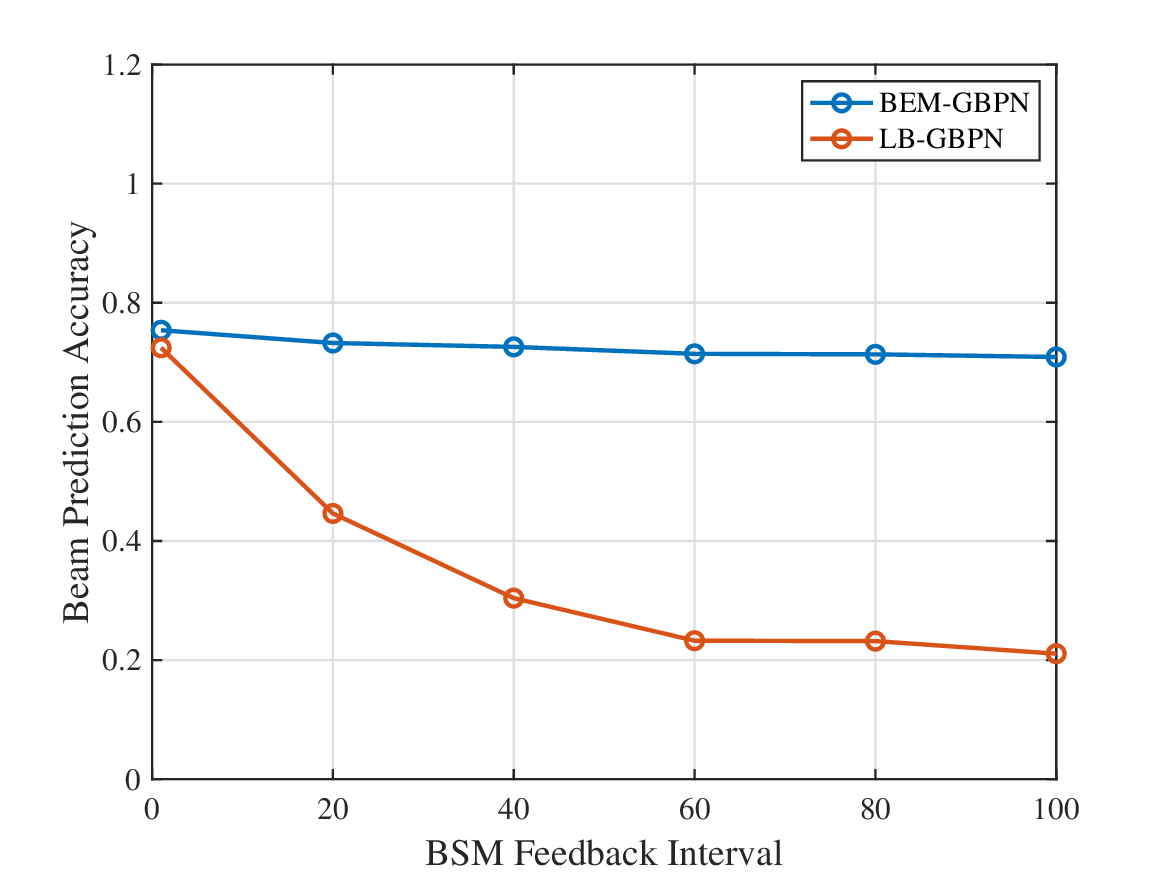}
    \caption{The beam prediction accuracy of the BEM-GBPN and LB-GBPN for BS1 versus the BSM feedback interval. } 
    \label{acc_bsm_fre}
    \vspace{-0.45cm}
\end{figure}
\subsubsection{Multi-User Beam Prediction}
We first evaluate the beam prediction accuracy of the BEM-GBPN. As shown in Fig. \ref{acc_nums_ue}, the BEM-GBPN achieves an average Top-$1$ \footnote{The Top-$k$ beam prediction represents selecting the best beam from the first $k$ predicted results of the network. } accuracy of $77.5\%$ and Top-$3$ accuracy of $99.7\%$ across $4$ BSs in single-user multi-scatterer scenarios. This performance validates both the efficacy of OOB modality perception and the network's capability for accurate user trajectory prediction. The subpar accuracy in the Top-$1$ beam prediction is due to vehicular dynamic pose variations, which induce antenna's position to deviate from the centroid of detected bounding boxes of these vehicles. Nevertheless, the predicted beams still point around the users, resulting in minimal communications quality degradation. In addition, as the number of users increases, the beam prediction accuracy drops significantly due to increasingly complex spatial distribution of mobile users and surrounding scatterers. However, even in a $4$-user scenario, each BS maintains a Top-$3$ accuracy above $98\%$, demonstrating the robustness of the proposed scheme for multi-user beam prediction.

Then, we compare the proposed BEM-GBPN with a classical compressed sensing (CS) based beam training method \cite{alkhateeb2015compressed}. In the CS-based approach, at time $(t+T_f)$, a small set of beams is designed to sense the channel, where the number of pilot beams $k_{cs} \leq N_T$. After performing sparse beam sweeping, the RSRP over the entire beam domain is reconstructed using the orthogonal matching pursuit (OMP) algorithm, and the beam with the highest RSRP is selected as the estimated optimal beam. For a fair comparison, at time $(t+T_f)$, the Top-$k_{cs}$ most probable beams predicted by the BEM-GBPN are selected for beam searching, and the beam with the highest received power is chosen as the optimal one. As shown in Fig. \ref{acc_nums_searches}, the beam alignment accuracy of both algorithms improves significantly with the increase in the number of beam searches. It is worth noting that the proposed BEM-GBPN achieves a beam alignment accuracy that is $69\%$ higher than that of the CS-based beam training method with zero beam sweeping overhead (i.e., $k_{cs}=1$). Moreover, with only $3$ beam searches, BEM-GBPN achieves $98\%$ beam alignment accuracy, reducing the beam sweeping overhead by $90\%$ compared with the CS-based method at the same accuracy level. These results demonstrate the advantages of OOB modality sensing in enhancing beam alignment performance. 

We further compare the proposed BEM-GBPN with some SOTA multi-modality aided beam prediction methods, including the bounding box based gain and beam prediction network (BB-GBPN) \cite{ahn2024sensinga}, the vision and partial beam sweeping (PBS) integration based gain and beam prediction network (VB-GBPN) \cite{jiang2022computer}, and the location and PBS integration based gain and beam prediction network (LB-GBPN) \cite{ardiantonugroho2025gpsaided}. Specifically, the BB-GBPN shares the same OMS-UIT module as the BEM-GBPN, while differing only in the data representation method. The bounding box based representation method needs to fix the number of PUOs, e.g., $10$, and the network input of the $u$-th UO is dentoed as $\boldsymbol{B}_{b,u}(t) = [Box'_{b,u}(t)^T, Box_{b,1}^T, ..., Box_{b,9}(t)^T]^T \in \mathbb{R}^{10 \times 4}$, in which the bounding box of the $u$-th UO is placed on the top row, and the other rows contain the bounding boxes of scatterers, i.e., $Box'_{b,u}(t) \in \boldsymbol{\mathcal{X}}^{UO}_b(t)$ and $\{Box_{b,1}(t), ..., Box_{b,9}(t)\} \in \boldsymbol{\mathcal{X}}^{Box}_b(t) \setminus Box'_{b,u}(t)$. In addition, the PBS refers to uniformly selecting a subset $\bm{\mathcal{F}}^p$ of beams from the codebook $\bm{\mathcal{F}}$ to perform a limited number of beam sweeping, i.e., $\bm{\mathcal{F}}^p=\{\boldsymbol{f}_1, \boldsymbol{f}_{1+\frac{N_T}{\gamma_p}}, ..., \boldsymbol{f}_{1+(\gamma_p-1)\frac{N_T}{\gamma_p}}\}$, where $\gamma_p=4$ denotes the number of beams in the $\bm{\mathcal{F}}^p$. Subsequently, the BB-GBPN adopts a four-layer multilayer perceptron (MLP) architecture, with $(128, 256, 512, 128)$ neurons, for the SFE. The VB-GBPN employs a ResNet-$18$ \cite{he2016deep} to extract features from each resized raw image, while another MLP architecture, with $(128, 256, 512, 64)$ neurons, is used to extract channel features from PBS. The two feature vectors are then concatenated into a $128$ dimensional embedding vector. In contrast, the LB-GBPN uses two MLP networks, each with $(128, 256, 512, 64)$ neurons, to extract features from location of users and PBS, respectively, and concatenates them into a $128$ dimensional embedding vector. All other components of BB-GBPN, VB-GBPN, and LB-GBPN are consistent with the BEM-GBPN. Furthermore, to evaluate the effectiveness of the MoE structure, we replace the MoE structure of the proposed BEM-GBPN with two MLPs, consisting of $(256,32,1)$ and $(256, 128, 64)$ neurons, for beamforming gain and optimal beam predictions, respectively. 
\begin{figure}[t]
    \centering
    \includegraphics[scale=0.4]{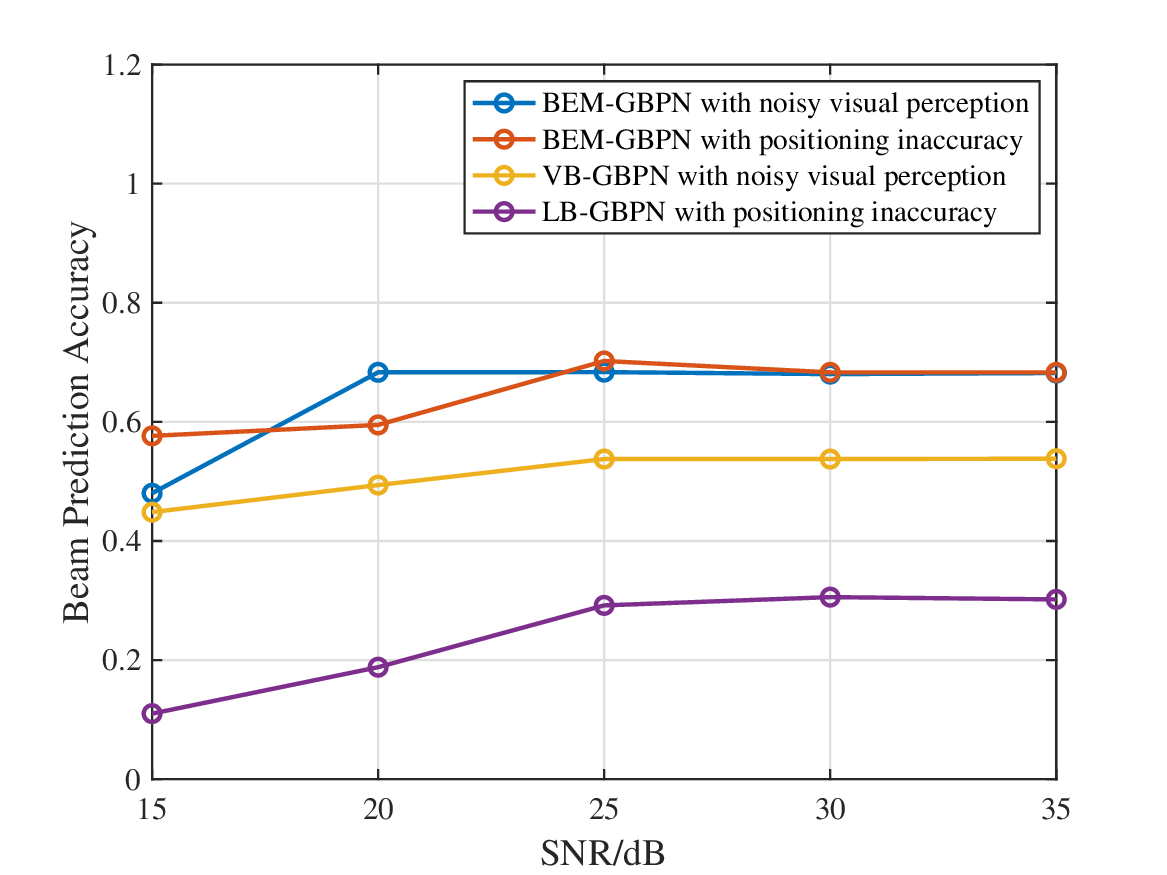}
    \caption{The beam prediction accuracy of the BEM-GBPN, VB-GBPN, and LB-GBPN for BS1 with noise, and the BSM feedback interval is fixed at $40$. } 
    \label{acc_snr}
    \vspace{-0.5cm}
\end{figure}

As shown in Fig. \ref{acc_epochs}, the proposed BEM-GBPN achieves an accurate Top-$1$ beam prediction performance, which is $45\%$ higher and $17.5\%$ higher than that of the BB-GBPN and the VB-GBPN, respectively. This indicates that the BEM-based representation method can more effectively represent the spatial distribution characteristics than the bounding box based representation method, since unbalanced numbers of available network inputs in the training dataset will lead to poor generalization in scenarios with variable numbers of users and scatterers. While in the VB-GBPN, raw images are directly used as the input without explicitly distinguishing the background, mobile users, and mobile scatterers, which increases the training complexity and degrades the prediction stability. In addition, the LB-GBPN achieves a performance comparable to that of the BEM-GBPN with accurate location feedback from the user and PBS, while the proposed BEM-GBPN eliminates the pilot overhead generated by the PBS. Moreover, the MoE architecture effectively enhances model performance by $6\%$ without significantly increasing model complexity. 
\begin{figure}[t]
    \centering
    \includegraphics[scale=0.4]{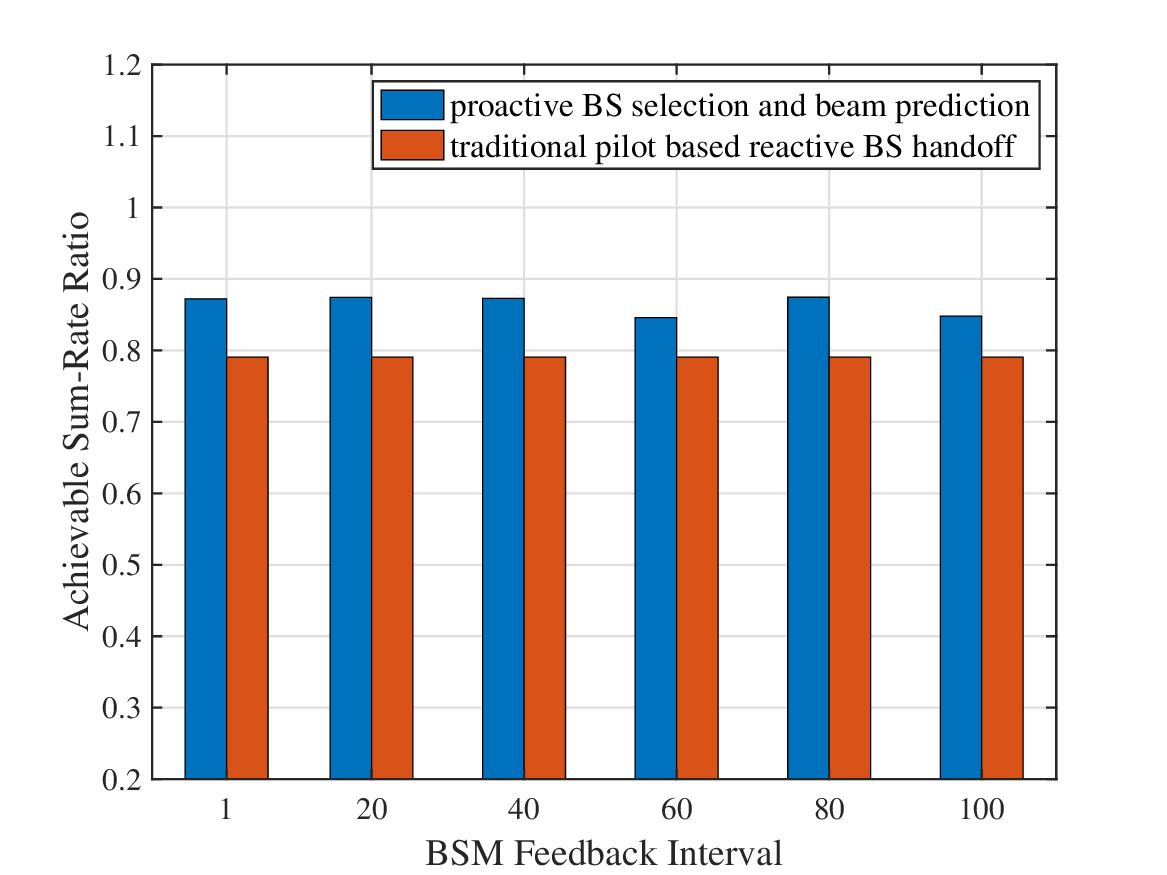}
    \caption{The ASRR of the proactive BS selection and beam switching approach and traditional pilot based reactive handoff versus the BSM feedback interval in a single-user multi-scatterer scenario.}
    \label{asrr_bsm_fre_react}
    \vspace{-0.5cm}
\end{figure}

In practical vehicular networks, interruptions in BSM feedback, inaccurate positioning, and low-quality visual perception are inevitable. In Fig. \ref{acc_bsm_fre}, we evaluate the impact of different BSM feedback intervals on beam prediction accuracy of the proposed BEM-GBPN and LB-GBPN. When BSM feedback is lost, the LB-GBPN continues to use outdated location sequences combined with real-time PBS features for beam prediction, causing a significant performance degradation as the BSM feedback interval increases. In contrast, the proposed OMS-UIT algorithm leverages visual sensing to continuously track users for the BEM-GBPN during BSM interruption periods, thereby substantially mitigating the instability caused by unreliable BSM feedback. Furthermore, we add noise to the OOB modalities, i.e., location and image, to emulate positioning inaccuracy and adverse visual perception conditions. Fig. \ref{acc_snr} shows the impact of positioning inaccuracy on different methods with a fixed BSM feedback interval of $40$ (i.e., $800$ ms). The SNR of the noisy locations can be expressed as $\mathrm{SNR}_{loc}=10\log_{10}\frac{x^2+y^2}{2\sigma_l^2}$, where $(x, y)$ represents the location of the vehicle, and $\sigma_l$ is the variance of the added AWGN \cite{lin2024multicameraa}. The performance of LB-GBPN exhibits the most severe degradation with the decrease of SNR, whereas the BEM-GBPN, benefiting from its visual object detection capability, can correct and compensate for positioning inaccuracies, thus achieving more robust and stable performance. Even at a better location SNR of $35$ dB, the performance of LB-GBPN is still $38\%$ lower than that of the BEM-GBPN. Finally, we added noise to the images to evaluate the robustness of visual perception. The image SNR can be expressed as  $\mathrm{SNR}_{img}=10\log_{10}\frac{\|\boldsymbol{I}\|_2^2}{|\boldsymbol{I}| \sigma_i^2}$, where $\boldsymbol{I}$ represents the image tensor, and $\sigma_i$ is the variance of the added AWGN \cite{lin2024multicameraa}. As shown in Fig. \ref{acc_snr}, the performance of BEM-GBPN degrades severely at a lower image SNR of $15$ dB, because the noise affects the front-end object detection in the input images. Nevertheless, the performance of BEM-GBPN still significantly outperforms VB-GBPN. These results demonstrate that the BEM-based representation enables the proposed BEM-GBPN to achieve higher robustness compared with the VB-GBPN, which directly uses raw images as the input.
\begin{figure}[t]
    \centering
    \includegraphics[scale=0.4]{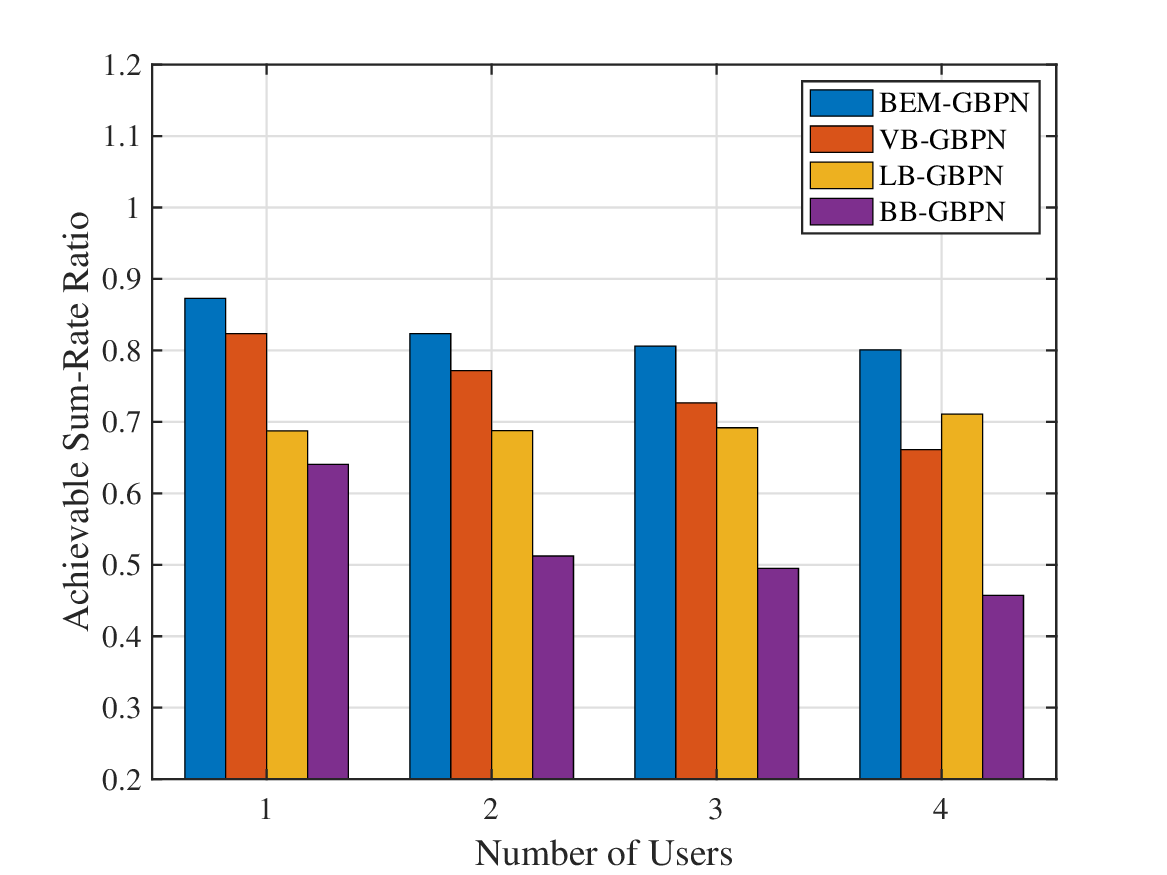}
    \caption{The ASRR of the proactive BS selection and beam switching approach based on BEM-GBPN, VB-GBPN, LB-GBPN, and BB-GBPN, versus the number of users, with the BSM feedback interval fixed at $40$. }
    \label{asrr_nums_ue}
    \vspace{-0.5cm}
\end{figure}
\begin{figure}[t]
    \centering
    \includegraphics[scale=0.4]{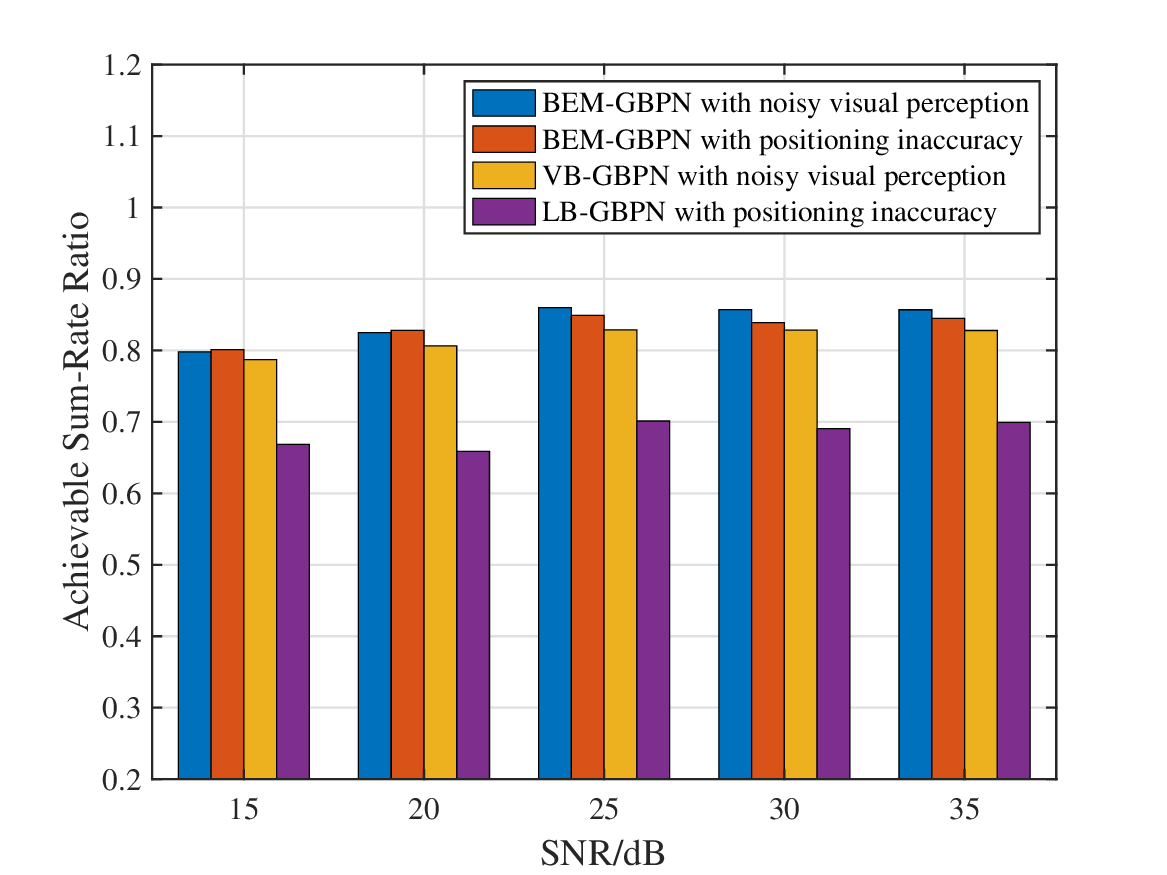}
    \caption{The ASRR of the proactive BS selection and beam switching approach based on BEM-GBPN, VB-GBPN, and LB-GBPN with noise, and the BSM feedback interval is fixed at $40$. }
    \label{asrr_snr}
    \vspace{-0.5cm}
\end{figure}
\subsubsection{Base Station Handoff and Beam Switching}
We adopt an achievable sum-rate ratio (ASRR), formulated as $\frac{\sum_{u\in \bm{\mathcal{U}}} R_u}{\sum_{u\in \bm{\mathcal{U}}} R^{opt}_u}$, to evaluate the quality of communication links during BS handoff, where $\sum_{u\in \bm{\mathcal{U}}} R_u$ represents the actual achievable sum-rate under the evaluated handoff scheme, and $\sum_{u\in \bm{\mathcal{U}}} R^{opt}_u$ refers to the optimal achievable sum-rate obtained with optimal mobility management, i.e., each user connects to the optimal BS and beam with the highest beamforming gain at every time slot. To fully evaluate the proactive BS selection and beam switching approach, we first compare the ASRR of the proposed approach with that of the traditional pilot based reactive handoff approach \cite{giordani2016multiconnectivity} in a single-user multi-scatterer scenario. The traditional pilot based reactive handoff is performed when the beamforming gain measured by the pilot of a BS is greater than that of the currently connected BS, such that the connection beam is selected via CS-based beam training utilizing $k_{cs}=5$ pilot beams from a codebook of size $64$. Assuming that the handoff and beam training incur a delay of $T_f=200$ ms, the proactive approach operates preemptively without incurring such latency. As shown in Fig. \ref{asrr_bsm_fre_react}, the BEM-GBPN based proactive BS selection and beam prediction achieves an ASRR of over $86\%$, which is $7\%$ higher than that of the traditional pilot based reactive BS handoff approach, highlighting the significant performance improvement of proactive handoff and precise BS selection based on beamforming gain prediction from the BEM-GBPN. In addition, with the increasing BSM feedback interval, the ASRR of the proposed scheme does not drop significantly due to visual tracking, which showcases the effectiveness of the OMS-UIT algorithm. Even with a BSM feedback interval of $100$ (i.e., $2000$ms), which is stringent in practical scenarios, the proposed BEM-GBPN based proactive BS selection and beam prediction can still achieve a transmission rate comparable to the optimal without any pilot overhead, thereby saving considerable spectrum resources. 

Moreover, to systematically demonstrate the performance of the proactive BS selection and beam switching approach in multi-user scenarios, we fix the BSM feedback interval at $40$ (i.e., $800$ ms), and the optimal BS and beam are predicted by the BEM-GBPN, BB-GBPN, VB-GBPN, and LB-GBPN, respectively. Fig. \ref{asrr_nums_ue} shows that, with an increasing number of users, the ASRR of approaches with visual sensing, i.e., BEM-GBPN, BB-GBPN, and VB-GBPN, drops significantly due to increasingly complex spatial distribution of mobile objects. Nevertheless, the BEM-GBPN still maintains an ASRR of $80\%$ in the $4$-user scenario, which is $13.9\%$ and $9\%$ higher than that of the VB-GBPN and LB-GBPN, respectively. Notably, the mobility management scheme with BB-GBPN suffers the most performance degradation with the increasing number of users, indicating that the vector-form bounding box struggles to represent variations in spatial characteristics and achieve sufficient environmental awareness.

Finally, we evaluate the impacts of positioning inaccuracy and adverse visual perception conditions on the algorithms' ASRR performance. The BSM feedback interval is fixed as $40$ (i.e., $800$ ms), and the noise level is configured in the same manner as described in Section IV.C.1). Fig. \ref{acc_snr} shows that inaccurate positioning significantly degrades the performance of both BEM-GBPN and LB-GBPN. Nevertheless, the LB-GBPN experiences the most severe performance degradation. Even at a better localization SNR of $35$ dB, its performance remains $14.5\%$ lower than that of the BEM-GBPN. Furthermore, when the image SNR is low, the OMS-UIT in the front-end of BEM-GBPN fails to detect some targets, leading to slightly inferior performance compared to the VB-GBPN, which operates directly on raw images. However, when the visual perception SNR exceeds $20$ dB, the performance improvement of the BEM-based representation becomes significant. Overall, compared with the VB-GBPN and the LB-GBPN, relying on PBS, the proposed BEM-GBPN achieves more robust performance under various noise conditions with zero beam sweeping overhead, demonstrating the superiority of the OMS-based mobility management scheme.
\begin{table}
\centering
\caption{Computational complexity comparison of the BEM-GBPN, BB-GBPN, VB-GBPN, and LB-GBPN.}
\label{complexity}
\begin{tabular}{l|cc}
\hline
\multirow{2}{*}{\makecell{\textbf{Method}}} & \multicolumn{2}{c}{\textbf{Computational Complexity}} \\
\cline{2-3} 
 & \makecell{\textbf{Space Complexity} \\ \textbf{(Params)}} & \makecell{\textbf{Time Complexity} \\ \textbf{(FLOPs)}} \\
\hline
\textbf{BEM-GBPN} & $1.85\times10^6$ & $1.39\times10^9$ \\
\hline
\textbf{BEM-GBPN w/o MoE} & $1.72\times10^6$ & $1.39\times10^9$ \\
\hline
\textbf{BB-GBPN} & $1.093\times10^6$ & $2.56\times10^7$ \\
\hline
\textbf{VB-GBPN} & $1.23\times10^7$ & $9.7\times10^{10}$ \\
\hline
\textbf{LB-GBPN} & $1.26\times10^6$ & $2.89\times10^7$ \\
\hline
\end{tabular}
\vspace{-0.5cm}
\end{table}
\subsubsection{Computational Complexity}
Here, we analyze the computational complexity of the proposed OMS-based mobility management scheme. Since only lightweight prediction results (e.g., user identities and predicted gains) need to be transmitted from BSs to the central unit, and the central unit merely sends lightweight control commands for proactive BS handoff and beam switching, the computational complexity of the OMS-based mobility management scheme is dominated by the OMS-UIT algorithm and the BEM-GBPN. Specifically, the user identification approach utilizes the YOLOv5-small (YOLOv5s) model, with $7.2\times10^6$ parameters and $1.65\times 10^{10}$ FLOPs, for object detection. Then, in the user tracking phase, the ByteTrack algorithm operates as a plug-in module, whose computational complexity is negligible. In addition, according to Table \ref{complexity}, the MoE architecture introduces a $7.5\%$ increase in space complexity, while significantly enhancing the performance of BEM-GBPN. The BEM-GBPN and VB-GBPN, both of which utilize the visual modality, exhibit relatively higher computational complexity. However, the BEM-GBPN reduces computational complexity by about $90\%$ compared with the VB-GBPN, owing to the efficient BEM-based representation. Although the LB-GBPN has lower complexity, it is more vulnerable to positioning errors and interruptions, and has higher beam sweeping overhead compared with the BEM-GBPN. The BB-GBPN has the lowest complexity, but its bounding box based representation leads to poor performance in both beam prediction and BS selection. Overall, the proposed OMS-based mobility management scheme with BEM-GBPN demonstrates the best trade-off among performance, computational complexity, and spectrum efficiency. 

We adopt a workstation with an 11-th Generation Intel CPU and an NVIDIA RTX 3090 GPU to implement the OMS-based mobility management scheme. During the user identification and tracking phase, the BSM feedback introduces negligible time overhead, and the OMS-UIT algorithm can achieve user identification within $4$ms. Note that once users are identified, they can be tracked seamlessly using the proposed scheme. Thus, the latency of the OMS-UIT algorithm will be even lower in practice. During the beamforming gain and optimal beam prediction phase, the inference time of the BEM-GBPN for one sample is approximately $15.2$ms under various settings of user numbers. Since the beamforming gain and optimal beam prediction for each user can be performed in parallel, the varying number of users will not significantly introduce additional latency. 
\subsubsection{Scalability Analysis}
To further evaluate the practicality of the proposed OMS-based mobility management scheme in ultra-dense vehicular networks, we analyze the coordination overhead and scalability of the distributed multi-BS cooperation scheme compared with conventional centralized approaches \cite{xu2023multiusera, lin2024multicameraa, ahn2024sensinga}. 

In the centralized architecture, all BSs must upload their raw multi-modal sensing data (e.g., RGB images and BSMs) to a central unit for joint processing. The required coordination backhaul throughput is
\begin{equation}\label{centralize_coordination}
BW_{cent} = B \cdot \frac{1}{T_s} \cdot (S_{img}+S_{aux}),
\end{equation}
where $B$ is the number of BS, $W_{cent}$ denotes the centralized coordination backhaul throughput per BS, $T_s=20$ms is the frame interval, and $S_{img}$ and $S_{aux}$ denote the per-frame sizes of the image and auxiliary metadata, respectively. 

In contrast, the proposed distributed architecture performs local inference at each BS and transmits only a few bytes of prediction results, i.e., the identity of detected users and the corresponding predicted beamforming gains, to the central unit. The coordination backhaul throughput can be expressed as
\begin{equation}\label{distributed_coordination}
BW_{dist} = B \cdot U_b \cdot \frac{1}{T_s} \cdot S_{pred},
\end{equation}
where $U_b$ denotes the number of users served by each BS, $W_{dist}$ denotes the distributed coordination backhaul throughput per BS, and $S_{pred}$ represents the payload size of a single user's prediction record. 

Given an image resolution of $480\times320$, the compressed image size is approximately $S_{pred}=60$KB, and auxiliary metadata is $S_{aux}=10$B. The prediction record is $S_{pred} = 16$B. Hence, the distributed scheme reduces coordination overhead by
\begin{equation}\label{distributed_coordination}
\frac{BW_{cent}}{BW_{dist}} \approx \frac{3750}{U_b}, 
\end{equation}
achieving more than $\frac{3750}{U_b}\times$ reduction in multi-BS coordination overhead. For an ultra-dense network, e.g., $B=100$ BSs, the centralized system requires $300$MB/s of backhaul throughput, whereas the proposed distributed design needs only $0.08$MB/s per user. In summary, the proposed distributed multi-BS coordination scheme maintains a low backhaul throughput requirement even in an ultra-dense vehicular network with hundreds of BS and users, demonstrating its significant advantage over the centralized scheme with respect to the scalability.
\section{Conclusion and Future Work}
In this paper, we proposed an out-of-band modality synergy based mobility management scheme to realize multi-user beam prediction and proactive BS selection, so as to eliminate pilot overhead and achieve efficient multi-BS coordination in multi-BS systems. Specifically, the OMS-UIT algorithm was proposed to initially identify and track users without any pilot overhead. Subsequently, the BEM-GBPN was proposed to predict the beamforming gains and optimal beams for multiple users. Finally, a proactive BS selection and beam switching approach based on beamforming gain and optimal beam predictions from all BSs was proposed to reduce delays and improve spectral efficiency significantly. Extensive simulation results demonstrated that the BEM-GBPN achieved significant performance improvements in both beamforming gain and beam prediction compared with existing methods. Additionally, the proposed proactive BS selection and beam switching approach enables a higher transmission rate than the traditional reactive BS handoff approach while saving considerable time and spectrum resources. In future research, adopting hybrid sensing frameworks by fusing vision with radar or LiDAR is expected to compensate for the limitations of visual perception in non-ideal environments (e.g., adverse weather and poor lighting conditions). Such multi-modal integration will further enhance communication performance and reliability.

\bibliographystyle{IEEEtran}
\bibliography{Refs}
\vspace{-1.5cm}
\begin{IEEEbiography}[{\includegraphics[width=1in,height=1.2in,clip,keepaspectratio]{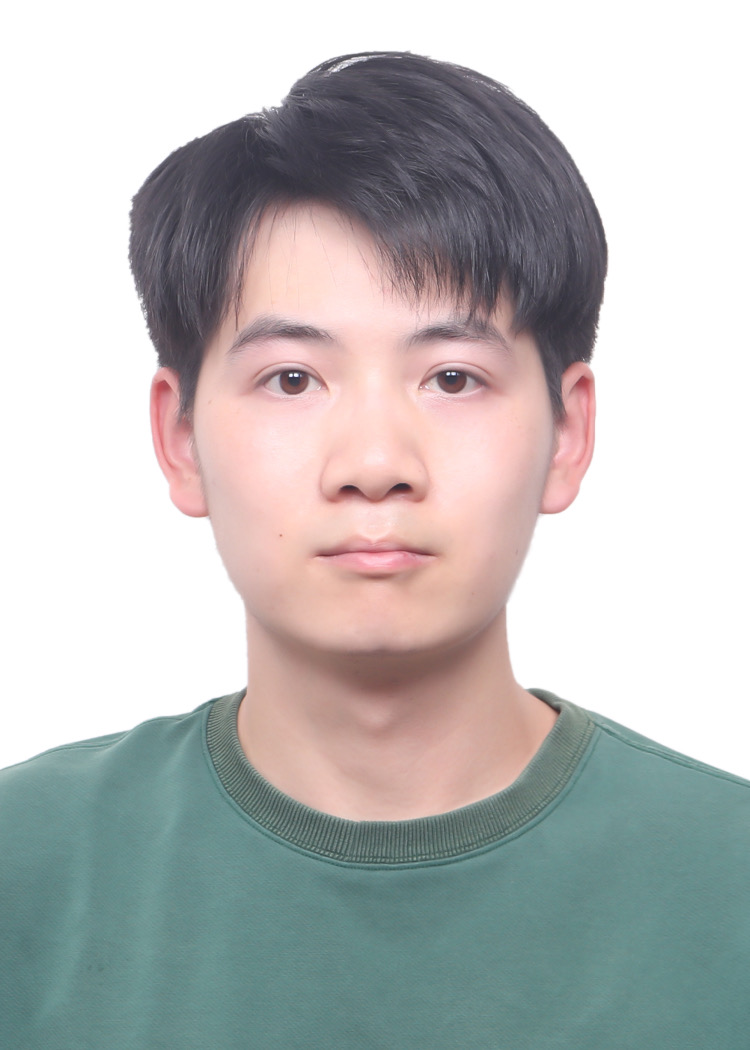}}]{Kehui Li} received his B.Eng. degree in Electrical Engineering from Tianjin University, Tianjin, China, in 2021, and his M.Sc. degree in Electrical and Computer Engineering from the University of Macau, Macao SAR, China, in 2024, respectively. He is currently working toward the Ph.D. degree in Electrical and Computer Engineering with the University of Macau, Macao SAR, China. His research interests include artificial intelligence-native communications, multimodal integrated sensing and communications (ISAC), and foundation models.
\end{IEEEbiography}
\vspace{-1.5cm}
\begin{IEEEbiography}[{\includegraphics[width=1in,height=1.2in,clip,keepaspectratio]{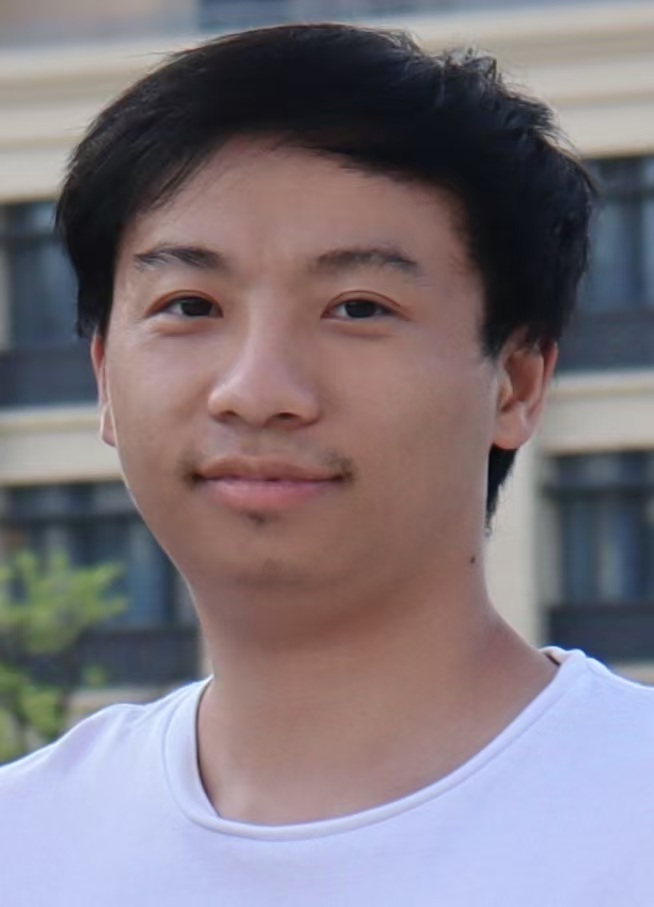}}]{Binggui Zhou} received his B.Eng. degree in Electrical Engineering from Jinan University, Zhuhai, China, in 2018, and his M.Sc. degree and Ph.D. degree in Electrical and Computer Engineering from the University of Macau, Macao SAR, China, in 2021 and 2024, respectively. He is currently a Postdoctoral Research Associate with the Department of Electrical and Electronic Engineering, Imperial College London, London, United Kingdom. His research interests include machine learning and data science as well as their applications in wireless communications and smart healthcare.
\end{IEEEbiography}
\vspace{-1.5cm}
\begin{IEEEbiography}[{\includegraphics[width=1in,height=1.2in,clip,keepaspectratio]{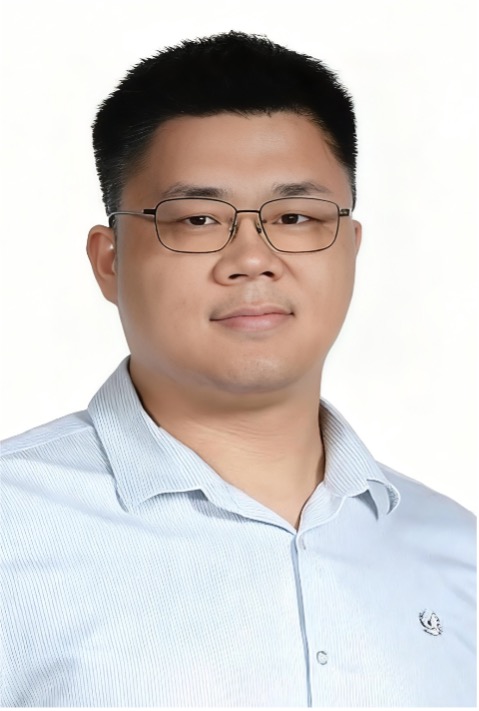}}]{Jiajia Guo} received the B.S. degree from the Nanjing University of Science and Technology, Nanjing, China, in 2016, the M.S. degree from the University of Science and Technology of China, Hefei, China, in 2019, and the Ph.D. degree in information and communications engineering from Southeast University, Nanjing, China, in 2023.  He has been serving as a Postdoctoral Fellow in State Key Laboratory of Internet of Things for Smart City, University of Macau since 2023. His research interests focus on AI-native air interfaces, massive MIMO, ISAC, and large AI models. His achievements were selected as one of the Top 10 Science and Technology Progress in the Information and Communication field for 2022 in China. Dr. Guo was the recipient of the First Prize from Natural Science of the Chinese Institute of Electronics and the 2023 Chinese Institute of Electronics Best Doctoral Thesis Award.
\end{IEEEbiography}
\vspace{-1.5cm}
\begin{IEEEbiography}[{\includegraphics[width=1in,height=1.2in,clip,keepaspectratio]{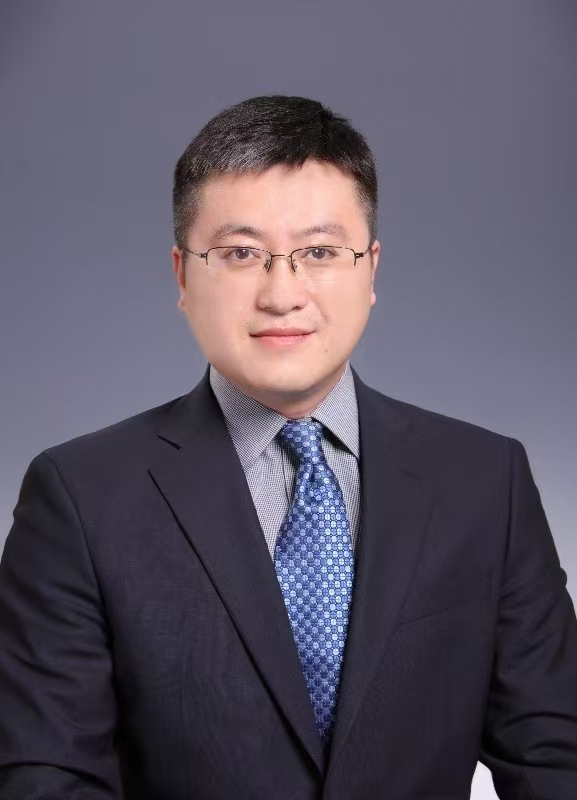}}]{Feifei Gao}  (Fellow, IEEE) received the B.Eng. degree from Xi'an Jiaotong University, Xi'an, China in 2002, the M.Sc. degree from McMaster University, Hamilton, ON, Canada in 2004, and the Ph.D. degree from National University of Singapore, Singapore in 2007. Since 2011, he joined the Department of Automation, Tsinghua University, Beijing, China, where he is currently a tenured full professor. 

Prof. Gao's research interests include signal processing for communications, array signal processing, convex optimizations, and artificial intelligence assisted communications. He has authored/coauthored more than 200 refereed IEEE journal papers and more than 150 IEEE conference proceeding papers that are cited more than 24000 times in Google Scholar. Prof. Gao has served as an Editor of IEEE Transactions on Wireless Communications, IEEE Journal of Selected Topics in Signal Processing (Lead Guest Editor), IEEE Transactions on Cognitive Communications and Networking, IEEE Signal Processing Letters (Senior Editor), IEEE Communications Letters (Senior Editor), IEEE Wireless Communications Letters, and China Communications. He has also served as the symposium co-chair for 2019 IEEE Conference on Communications (ICC), 2018 IEEE Vehicular Technology Conference Spring (VTC), 2015 IEEE Conference on Communications (ICC), 2014 IEEE Global Communications Conference (GLOBECOM), 2014 IEEE Vehicular Technology Conference Fall (VTC), as well as Technical Committee Members for more than 50 IEEE conferences.
\end{IEEEbiography}
\vspace{-1.5cm}
\begin{IEEEbiography}[{\includegraphics[width=1in,height=1.2in,clip,keepaspectratio]{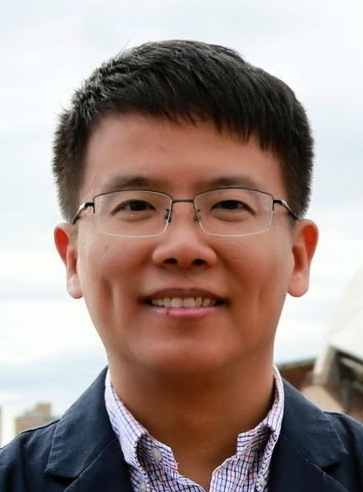}}]{Guanghua Yang} (Senior Member, IEEE) received his Ph.D. degree in electrical and electronic engineering from the University of Hong Kong in 2006. From 2006 to 2013, he served as post-doctoral fellow, research associate at the University of Hong Kong. Since April 2017, he has been with Jinan University, where he is currently a Full Professor in the School of Intelligent Systems Science and Engineering. His research interests are in the general areas of communications and networking.
\end{IEEEbiography}
\vspace{-1.5cm}
\begin{IEEEbiography}[{\includegraphics[width=1in,height=1.2in,clip,keepaspectratio]{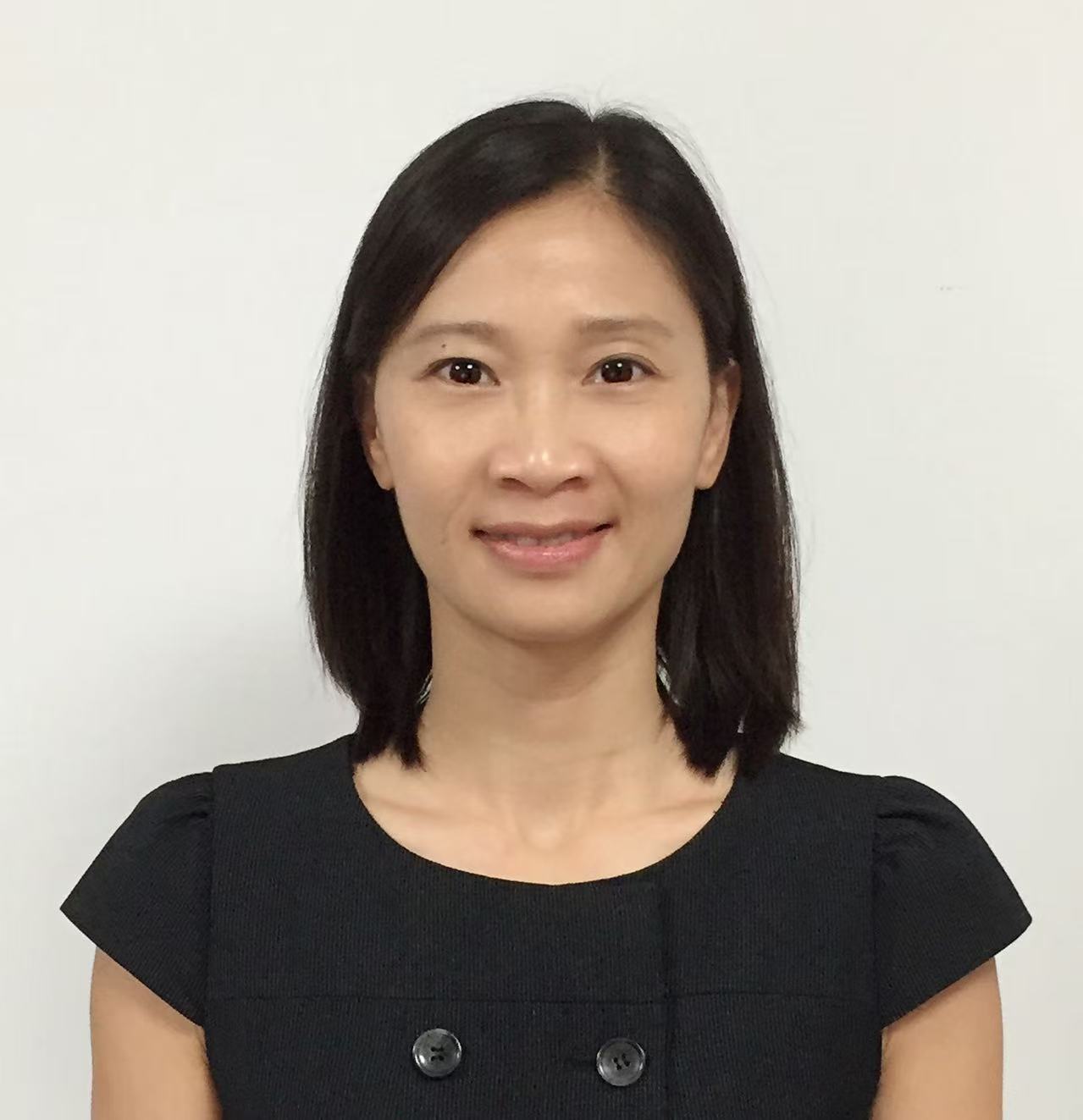}}]{Shaodan Ma} (Senior Member, IEEE) received the double Bachelor’s degrees in science and economics and the M.Eng. degree in electronic engineering from Nankai University, Tianjin, China, in 1999 and 2002, respectively, and the Ph.D. degree in electrical and electronic engineering from The University of Hong Kong, Hong Kong, in 2006. From 2006 to 2011, she was a post-doctoral fellow at The University of Hong Kong. Since August 2011, she has been with the University of Macau, where she is currently a Professor. Her research interests include array signal processing, transceiver design, localization, integrated sensing and communication, mmWave/THz communications, massive MIMO, and machine learning for communications. She was a symposium Co-Chair for various conferences including IEEE VTC2024-Spring, IEEE ICC 2021, 2019 \& 2016, IEEE GLOBECOM 2016, IEEE/CIC ICCC 2019, etc. She is an IEEE ComSoc Distinguished Lecturer in 2024-2025 and has served as an Editor for {\scshape IEEE Transactions on Cognitive Communications and Networking} (2025-present), {\scshape IEEE Wireless Communications} (2024-present), {\scshape IEEE Communications Letters} (2023), {\scshape Journal of Communications and Information Networks} (2021-present), {\scshape IEEE Transactions on Wireless Communications} (2018-2023), {\scshape IEEE Transactions on Communications} (2018-2023), and {\scshape IEEE Wireless Communications Letters} (2017-2022).
\end{IEEEbiography}

\end{document}